\documentclass[a4paper,10pt,twoside]{cpc-hepnp}

\usepackage{multicol}
\usepackage{graphicx}
\usepackage{booktabs}
\usepackage{amssymb,bm,mathrsfs,bbm,amscd}
\usepackage[tbtags]{amsmath}
\usepackage{lastpage}
\usepackage{CJK}
\usepackage[usenames]{color}
\usepackage{xcolor}
\RequirePackage{multirow}
\RequirePackage{slashed}

\begin{document}
\begin{CJK*}{GBK}{song}

\fancyhead[c]{\small Chinese Physics C Vol. 42, No. 2 (2018) 023109} \fancyfoot[C]{\small -\thepage}

\footnotetext[0]{Received 31 October 2017}

\title{A data-driven approach to $\pi^{0}, \eta$ and $\eta^{\prime}$ single and double Dalitz decays}

\author{%
      Rafel Escribano$^{a,b;1)}$\email{rescriba@ifae.es}%
\quad Sergi Gonz\`{a}lez-Sol\'is$^{b,c;2)}$\email{sgonzalez@itp.ac.cn}%
}
\maketitle

\address{%
$^a$ Grup de F\'{\i}sica Te\`orica, 
Departament de F\'{\i}sica, 
Universitat Aut\`onoma de Barcelona, 
E-08193 Bellaterra (Barcelona), Spain\\
$^b$ Institut de F\'{\i}sica d'Altes Energies (IFAE), 
The Barcelona Institute of Science and Technology (BIST), 
Campus UAB,
E-08193 Bellaterra (Barcelona), Spain\\
$^c$ CAS Key Laboratory of Theoretical Physics, Institute of Theoretical Physics, Chinese Academy of Sciences, Beijing 100190, China\\
}

\begin{abstract}
The dilepton invariant mass spectra and integrated branching ratios of the single and double Dalitz decays 
$\mathcal{P}\to\ell^{+}\ell^{-}\gamma$ and $\mathcal{P}\to\ell^{+}\ell^{-}\ell^{+}\ell^{-}$ 
($\mathcal{P}=\pi^{0}, \eta, \eta^{\prime}$; $\ell=e$ or $\mu$) are predicted by means of a 
data-driven approach based on the use of rational approximants applied to
 $\pi^{0}, \eta$ and $\eta^{\prime}$ transition form factor experimental data in the space-like region.
\end{abstract}


\begin{pacs}
13.20.Cz, 12.15.Hh, 12.40.Vv\qquad{\bf{DOI:}} 10.1088/1674-1137/42/2/023109
\end{pacs}

\footnotetext[0]{\hspace*{-3mm}\raisebox{0.3ex}{$\scriptstyle\copyright$}2013
Chinese Physical Society and the Institute of High Energy Physics
of the Chinese Academy of Sciences and the Institute
of Modern Physics of the Chinese Academy of Sciences and IOP Publishing Ltd}%

\begin{multicols}{2}

\section{Introduction}

Anomalous decays of the neutral pseudoscalar mesons $\mathcal{P}$ $(\mathcal{P}=\pi^{0},\eta,\eta^{\prime})$ 
are driven through the chiral anomaly of QCD.
Of historical importance is the process $\mathcal{P}\to\gamma\gamma$.
Apart from being the main decay channel of the $\pi^{0}$ and the $\eta$, 
its experimental discovery confirmed, for the first time, the existence of anomalies.
In this case, the two final state photons are real and the transition form factor (TFF) 
encoding the effects of the strong interactions of the decaying meson is predicted to be simply a constant, 
the value of the axial anomaly in the chiral and large-$N_{c}$ limits of QCD, 
$F_{\pi^{0}\gamma\gamma}(0)=1/(4\pi^{2}F_{\pi})$ in the case of neutral pion, where $F_{\pi}\simeq 92$ MeV is the pion decay constant. 
However, if one of the two photons is virtual, 
the corresponding TFF is no longer a constant but a function of the transferred momentum to the virtual photon 
$F_{\mathcal{P}\gamma\gamma^{\ast}}(q^{2})$,
whereas when both photons are virtual the TFF depends on both photon virtualities and is represented by 
$F_{\mathcal{P}\gamma^{\ast}\gamma^{\ast}}$ $(q_{1}^{2},q_{2}^{2})$. 
A single Dalitz decay occurs through the single-virtual TFF after the conversion of the virtual photon into a lepton pair,
while double Dalitz decays proceed with the TFF of double virtuality involving two dilepton pairs in the final state.
Dalitz decays are hence attractive processes to improve our knowledge of the TFFs of the 
$\mathcal{P}\gamma^{(\ast)}\gamma^{\ast}$ vertices. 
This is the main motivation of this work, together with predicting the invariant mass spectra and the branching ratios $(\mathcal{BR})$
of the decays $\mathcal{P}\to\ell^{+}\ell^{-}\gamma$ and $\mathcal{P}\to\ell^{+}\ell^{-}\ell^{+}\ell^{-}$,
with $\mathcal{P}=\pi^{0}, \eta, \eta^{\prime}$ and $\ell=e$ or $\mu$.

From the experimental side, the current status is the following.
The PDG reported value for the branching ratio of the decay $\pi^{0}\to e^{+}e^{-}\gamma$ 
is $(1.174\pm 0.035)\%$ \cite{lab1},
which is obtained from the PDG fits of the ratio
$\Gamma(\pi^{0}\to e^{+}e^{-}\gamma)/\Gamma(\pi^{0}\to\gamma\gamma)=(1.188\pm 0.035)\%$
(the latest measurement of this ratio, $(1.140\pm 0.041)\%$, was performed by ALEPH in 2008 \cite{lab2})
and $B(\pi^{0}\to\gamma\gamma)=(98.823\pm 0.034)\%$.
It is worth commenting that this is the second most important decay mode of the $\pi^{0}$.
The branching ratio for the decay  $\eta\to e^{+}e^{-}\gamma$ has  recently been measured by the A2 Collaboration at MAMI
 \cite{lab3}, $(6.6\pm 0.4\pm 0.4)\times 10^{-3}$ (see also the most recent result in Ref.~\cite{lab4}),
and the CELSIUS/WASA Collaboration \cite{lab5}, $(7.8\pm 0.5\pm 0.8)\times 10^{-3}$, 
whilst the PDG quoted value is $(6.9\pm0.4)\times 10^{-3}$ \cite{lab1},
in accordance with the result, $(6.72\pm 0.07\pm 0.31)\times 10^{-3}$,
from the WASA@COSY Collaboration \cite{lab6}.
The decay $\eta\to\mu^{+}\mu^{-}\gamma$ has been studied by the NA60 Collaboration at CERN SPS \cite{lab7},
though they do not provide a value for the branching ratio.
The PDG fit reports the value $(3.1\pm0.4)\times 10^{-4}$ \cite{lab1}.
The branching fraction for the decay $\eta^{\prime}\to e^{+}e^{-}\gamma$ has recently been  measured for the first time
by the BESIII Collaboration
\cite{lab8}, obtaining a value of $(4.69\pm 0.20\pm 0.23)\times 10^{-4}$.
To end, the decay $\eta^{\prime}\to\mu^{+}\mu^{-}\gamma$ was measured long ago at the SERPUKHOV-134 experiment with a value of
$(1.08\pm 0.27)\times 10^{-4}$ \cite{lab9}.
Regarding the double Dalitz decays, the KTeV Collaboration measured the branching ratio of the decay 
$\pi^{0}\to e^{+}e^{-}e^{+}e^{-}$, $(3.46\pm 0.19)\times 10^{-5}$ \cite{lab10},
thus averaging the PDG result to $(3.38\pm0.16)\times 10^{-5}$ \cite{lab1}.
The KLOE Collaboration reported the first experimental measurement of $\eta\to e^{+}e^{-}e^{+}e^{-}$ \cite{lab11},
$(2.4\pm0.2\pm0.1)\times10^{-5}$,
which is in agreement with the result, $(3.2\pm 0.9\pm 0.5)\times 10^{-5}$, provided by the WASA@COSY Collaboration
\cite{lab6}.
In Ref.~\cite{lab5}, upper bounds for the branching ratios of $\eta\to\mu^{+}\mu^{-}\mu^{+}\mu^{-}$, 
$<3.6\times 10^{-4}$, and $\eta\to e^{+}e^{-}\mu^{+}\mu^{-}$, $<1.6\times 10^{-4}$, both at $90\%$ CL, are reported.
Finally, no experimental evidence for $\eta^{\prime}\to e^{+}e^{-}e^{+}e^{-}$, $\eta^{\prime}\to \mu^{+}\mu^{-}\mu^{+}\mu^{-}$
and $\eta^{\prime}\to e^{+}e^{-}\mu^{+}\mu^{-}$ exists.

On the theory side, the effort is focused on encoding the QCD dynamical effects in the anomalous 
$\mathcal{P}\gamma^{(\ast)}\gamma^{\ast}$ vertices through the corresponding TFF functions\footnote{The
matrix element of the $\mathcal{P}\gamma^{(\ast)}\gamma^{(\ast)}$ transition is given for real or virtual photons by
$\mathcal{M}_{\mathcal{P}\to\gamma^{(\ast)}\gamma^{(\ast)}}=
i e^2\epsilon_{\mu\nu\alpha\beta}q_1^\mu q_2^\nu\varepsilon_1^\alpha\varepsilon_2^\beta
F_{\mathcal{P}\gamma^{(\ast)}\gamma^{(\ast)}}(q_1^2,q_2^2)$,
where $q_1$ and $q_2$ are the four-momenta of the photons, $\varepsilon_1$ and $\varepsilon_2$ are their polarization vectors,
and $F_{\mathcal{P}\gamma^{(\ast)}\gamma^{(\ast)}}(q_1^2,q_2^2)$ is the TFF function \cite{lab70}.}.
The exact momentum dependence of these TFFs over the whole energy region is not known;
we only have theoretical predictions from chiral perturbation theory and perturbative QCD (ChPT, pQCD),
thus constraining the low- and space-like large-momentum transfer regions, respectively. 
The TFF at zero-momentum transfer can be inferred either from the measured two-photon partial width,
\begin{equation}
|F_{\mathcal{P}\gamma\gamma}(0)|^{2}=
\frac{64\pi}{(4\pi\alpha)^{2}}\frac{\Gamma(\mathcal{P}\rightarrow\gamma\gamma)}{M_{\mathcal{P}}^{3}}\ ,
\label{Pgammagamma}
\end{equation}
or the prediction from the axial anomaly in the chiral and large-$N_{c}$ limits of QCD, as mentioned before,
while the asymptotic behaviour of the TFF at $Q^2\equiv -q^{2}\to\infty$ should exhibit the right falloff as $1/Q^{2}$
\cite{lab12}\footnote{Perturbative QCD
predicts $\lim_{Q^2\to\infty}Q^2 F_{\pi^{0}\gamma\gamma^{\ast}}(Q^2)=2F_{\pi}$.
Alternative values to this result exist, see for instance Refs.~\cite{lab13,lab14}, 
though they seem to be disfavoured, as pointed out in Refs.~\cite{lab15,lab16}.
For $\eta$ and $\eta^{\prime}$, see the asymptotic values obtained in Refs.~\cite{lab17,lab18}.}.
Furthermore, the operator product expansion (OPE) predicts the behaviour of the double-virtual TFF in the limit
$Q_1^2=Q_2^2\equiv Q^2\to\infty$ to be the same as for the single one, that is, $1/Q^2$ \cite{lab19}\footnote{The
OPE predicts for the case of the pion $\lim_{Q^2\to\infty}Q^2 F_{\pi^{0}\gamma^{\ast}\gamma^{\ast}}(Q^2,Q^2)=2F_{\pi}/3$.}.
For the intermediate-momentum transfer region,
the most common parameterisation of the TFF, widely used by experimental analyses,
is provided by the vector meson dominance model (VMD).
The dispersive representation of the TFF in terms of $q^{2}$, where $q^{2}$ is the photon virtuality in the time-like momentum region, 
can be written as
\begin{equation}
F_{\mathcal{P}\gamma\gamma^{\ast}}(q^{2})=\int_{s_0}^{\infty}ds\frac{\rho(s)}{s-q^{2}-i\epsilon}\ ,
\end{equation}
where $s_0$ is the threshold for the physical intermediate states imposed by unitarity and 
$\rho(s)=\mathrm{Im}F_{\mathcal{P}\gamma\gamma^{\ast}}(s)/\pi$ is the associated spectral function.
To approximate this intermediate-energy part of the spectral function, one usually employs one or more single-particle states.
As an illustration, the contribution to the spectral function of a narrow-width resonance of mass $M_{\rm{eff}}$ reduces to 
$\rho(s)\propto\delta(s-M^{2}_{\rm{eff}})$, which yields
\begin{equation}
F_{\mathcal{P}\gamma\gamma^{\ast}}(q^{2})=\frac{F_{\mathcal{P}\gamma\gamma}(0)}{1-q^2/\Lambda^2}\ ,
\label{poleFF}
\end{equation}
where $F_{\mathcal{P}\gamma\gamma}(0)$ serves as a normalisation constant and
$\Lambda(=M_{\rm{eff}})$ is a real parameter which fixes the position of the resonance pole on the real axis.
However, the simple and successful single-pole approximation given in Eq.~(\ref{poleFF}) breaks down for $q^{2}=\Lambda^{2}$.
One may cure this limitation by taking into account resonant finite-width effects as proposed by Landsberg in Ref.~\cite{lab70}
when considering the transitions $\mathcal{P}\to\ell^{+}\ell^{-}\gamma$ in a VMD framework.
According to this model, these transitions occur through the exchange of the lowest-lying $\rho$, $\omega$ and $\phi$ vector resonances
and their contributions to the TFF are written as
\begin{eqnarray}
\label{VMD}
\widetilde F_{\mathcal{P}\gamma\gamma^{\ast}}(q^{2})&=&
\left(\sum_{V=\rho,\omega,\phi}\frac{g_{V\mathcal{P}\gamma}}{2g_{V\gamma}}\right)^{-1}\\
&&\times\sum_{V=\rho,\omega,\phi}\frac{g_{V\mathcal{P}\gamma}}{2g_{V\gamma}}
\frac{M_{V}^{2}}{M_{V}^{2}-q^{2}-iM_{V}\Gamma_{V}(q^{2})}\ ,
\nonumber
\end{eqnarray}
where $\widetilde F_{\mathcal{P}\gamma\gamma^{\ast}}(q^{2})=
F_{\mathcal{P}\gamma\gamma^{\ast}}(q^{2})/F_{\mathcal{P}\gamma\gamma}(0)$
is defined as the normalised TFF, 
$g_{V\mathcal{P}\gamma}$ and $g_{V\gamma}$ are the $V\mathcal{P}\gamma$ and $V\gamma$ couplings, respectively,
$M_V$ the vector masses, and $\Gamma_{V}(q^{2})$ the energy-dependent widths.

Despite the notorious success of VMD in describing lots of phenomena at low and intermediate $q^2$,
particularly useful for the decays we consider in this work,
this model can be seen as a first step in a systematic approximation.
Pad\'e approximants are used to go beyond VMD in a simple manner, also incorporating information from higher energies,
allowing an improved determination of the low-energy constants relative to other methods  \cite{lab20}.
For this reason, we make use in our study of the work in Refs.~\cite{lab17,lab21},
where all current measurements of the space-like TFFs $\gamma^{\ast}\gamma\to\mathcal{P}$
\cite{lab22,lab23,lab24,lab25,lab26,lab27},
produced in the reactions $e^{+}e^{-}\to e^{+}e^{-}\mathcal{P}$,
have been accommodated in nice agreement with experimental data using these rational approximants.
We benefit from these parameterisations valid in the space-like region to predict the transitions 
$\mathcal{P}\gamma^{(\ast)}\gamma^{\ast}$ in the time-like region for the Dalitz decays we are interested in.
Despite some shortcomings Pad\'{e} approximants have in entirely describing the TFF analytical structure in the time-like region,
our primary aim is to achieve reasonable results for these decays which might serve as a guideline for the experimental collaborations.
Different parameterizations existing in the literature are based on interpolation formulas \cite{lab28,lab29},
resonance chiral theory \cite{lab30,lab31}
and dispersive techniques \cite{lab32,lab33}, among others
\cite{lab34,lab35,lab36,lab37,lab38,lab39,lab40,lab41,lab42,lab43}.

This paper is organised as follows.
In Section {\ref{FormFactor}}, we introduce our description of the $\pi^0$, $\eta$ and $\eta^\prime$ transition form factors
using the mathematical method of Pad\'{e} approximants. 
Sections {\ref{Single}} and {\ref{Double}} are devoted to the analysis of single and double Dalitz decays, respectively,
and predictions for several invariant mass spectra and branching ratios are given.
Finally, in Section {\ref{Conclusions}}, we present our conclusions.

\section{Transition form factors}

\label{FormFactor}
The usefulness of Pad\'{e} approximants (PAs) as fitting functions for different form factors has been illustrated,
for example, in Refs.~\cite{lab17,lab18,lab20,lab21,lab44,lab45}.
It is not our purpose to provide here formal details of the method but rather to cover some important aspects for consistency.
The PAs to a given function are ratios of two polynomials (with degree $L$ and $M$, respectively)\footnote{Without
any loss of generality, we take $b_0=1$ for definiteness.},
\begin{eqnarray}
\label{Pade}
P^{L}_{M}(q^{2})&=&\frac{\sum_{j=0}^{L}a_{j}(q^{2})^{j}}{\sum_{k=0}^{M}b_{k}(q^{2})^{k}}\\
&=&\frac{a_{0}+a_{1}q^{2}+\cdots+a_{L}(q^{2})^{L}}{1+b_{1}q^{2}+\cdots+b_{M}(q^{2})^{M}}\ ,\nonumber
\end{eqnarray}
constructed such that the Taylor expansion around the origin exactly coincides with that of $f(q^2)$ up to the highest possible order,
i.e., ${\mathcal O}(q^2)^{L+M+1}$.
We would like to point out that the previous VMD ansatz for the form factor, Eq.~(\ref{poleFF}),
can be viewed as the first element in a sequence of PAs which can be constructed in a systematic way.
By considering higher-order terms in the sequence, one may be able to describe the experimental space-like data
with an increasing level of accuracy.
The important difference with respect to the traditional VMD approach is that, as a Pad\'{e} sequence, 
the approximation is well-defined and can be systematically improved.
Although polynomial fitting is more common, in general, rational approximants are able to approximate the original function 
in a much broader range in momentum than a polynomial \cite{lab46}.
Once a Pad\'e sequence is chosen, the highest element belonging to that particular sequence will be fixed by experimental data.
To be specific, the sequence will be stopped as soon as the additional coefficients of the next order PA are statistically compatible with zero.

Despite the success of PAs as fitting functions for space-like TFFs, some important remarks are in order.
First, there is no {\emph{a priori}} mathematical proof ensuring the convergence of a Pad\'e sequence to the unknown TFF function,
though a pattern of convergence may be inferred from the data analysis {\emph{a posteriori}}\footnote{For
a detailed discussion of Pad\'e convergence applied to form factors, see for instance Ref.~\cite{lab20}.}.
For instance, the excellent performance of PAs in Ref.~\cite{lab18} (see Figs. 2 and 7 there)
seems to indicate that the convergence of the $\eta$ TFF normalisation and low-energy constants is assured
(see also Ref.~\cite{lab45} for the $\eta^{\prime}$ case).
Second, unlike the space-like TFF data analyses \cite{lab17,lab21}, 
one should not expect to reproduce the time-like TFF data,
since a Pad\'e approximant contains only isolated poles and cannot reproduce a time-like cut\footnote{The TFF function
is unknown but expected to be analytical in the entire $q^2$-complex plane, except for a branch cut along the real axis
for $q^2\ge 4M_\pi^2$.}.
However, if this right-hand cut is approximated by one or more single-particle states in the form of one or several narrow-width resonances,
as stated before, then the Pad\'e method may be still used up to the first resonance pole, indeed,
up to neighbourhoods of the pole.
A detailed explanation of the mathematical reasons for this being the case, upon discussion, can be found
in Section II of Ref.~\cite{lab45}\footnote{These reasons
are based on the fact that the imaginary part of the TFF at the $\pi\pi$ threshold and beyond is smooth
and thus acceptable up to the first resonance pole where the threshold expansion in powers of the pion momentum would break down.}.
Other approaches, such as the so-called $z$-expasion, incorporate {\emph{ab initio}} the unitary cut and are shown to be convenient
for describing form factors \cite{lab72}.
The size of the region which is affected by the presence of the pole, a disk of radius $\varepsilon$,
is not known but, as we will see later, may be deduced,
thus fixing the range of application of the PAs for time-like data.
Third and last, the poles found in the PAs fitting the TFFs cannot be directly associated with physical resonance poles
in the second Riemann sheet of the complex plane.
These, in turn, may be obtained following the prescriptions of Refs.~\cite{lab47,lab48,lab49},
which is beyond the scope of the present work.

We would like to emphasise that the use of PAs as fitting functions for some set of experimental data can be viewed
as an effective mathematical method which intrinsically contains relevant physical information of the function
represented by the data set. 
In this work, we benefit from the findings of Refs.~\cite{lab17,lab21},
where the $\pi^0$, $\eta$ and $\eta^\prime$ TFFs were fixed in the space-like region from the analysis of the intermediate process 
$\gamma\gamma^{*}\to\mathcal{P}$ by several experimental collaborations, to predict the time-like region of the same TFFs 
needed for the description of the reaction $\mathcal{P}\to\gamma\gamma^{*}$
and therefore for the single and double Dalitz decays studied here.
The extrapolated version of the TFFs used in this analysis, from the space-like region to the time-like one, are discussed case by case
in the following.

\subsection{$\pi^{0}\to\gamma\gamma^{*}$}
\begin{figure*}
\centering
\includegraphics[width=0.5\textwidth]{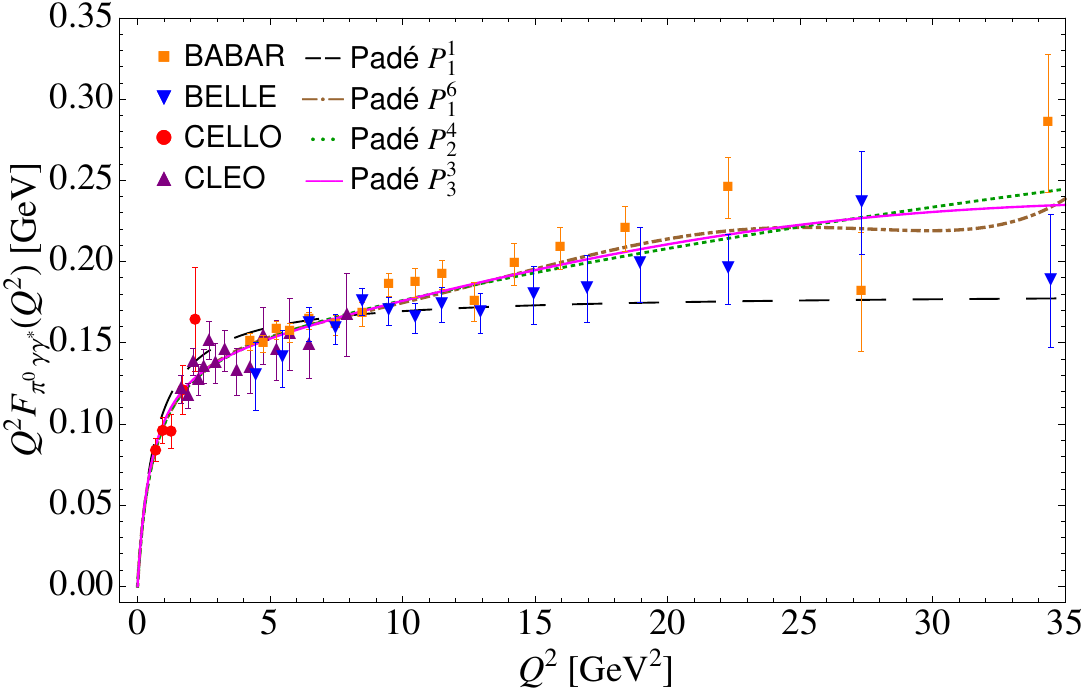}\includegraphics[width=0.5\textwidth]{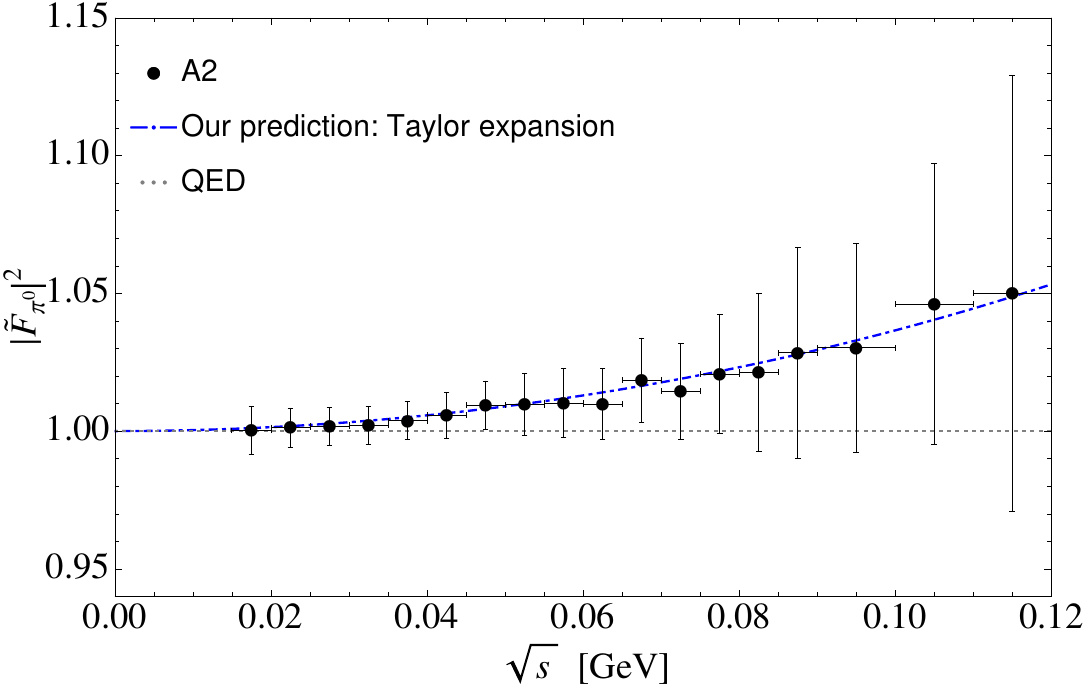}
\caption{Representation of the $\pi^{0}$ TFF as obtained from fits to the space-like experimental data using Pad\'{e} approximants \cite{lab21}
(a) and our prediction for its modulus square normalised time-like counterpart, $\widetilde F_{\pi^{0}\gamma\gamma^{\ast}}(q^{2})$,
as a function of the invariant dilelectron mass, $\sqrt{s}\equiv m_{ee}$ (b).
For the $\pi^{0}$ time-like TFF, the prediction coming from the Taylor expansion in Eq.~(\ref{Taylor}) (dot-dashed blue line)
is compared to the experimental data from $\pi^{0}\to e^{+}e^{-}\gamma$ \cite{lab71} (black circles).
The QED prediction (dotted grey line) is also displayed.}
\label{TFFpion}
\end{figure*}

Given the small phase-space available in the $\pi^{0}\to e^{+}e^{-}\gamma$ transition, $4m_{\ell}^2\le q^2\le M_\pi^2$,
the $\pi^0$ TFF can be expressed in terms of its Taylor expansion\footnote{The
form of the Taylor expansion is written in this way to make the slope and curvature parameters non-dimensional.},
\begin{equation}
F_{\pi^{0}\gamma\gamma^{*}}(q^{2})=
F_{\pi^{0}\gamma\gamma}(0)\left(1+b_{\pi}\frac{q^{2}}{M_{\pi}^{2}}+c_{\pi}\frac{q^{4}}{M_{\pi}^{4}}+\cdots\right),
\label{Taylor}
\end{equation}
where $F_{\pi^{0}\gamma\gamma}(0)$ is fixed from Eq.~(\ref{Pgammagamma})
and the values of the low-energy constants (LECs), slope and curvature, $b_\pi$ and $c_\pi$, respectively,
are borrowed from Eqs.~(12,13) in Ref.~\cite{lab21}\footnote{In 
Ref.~\cite{lab21}, the sign of the slope parameter in the Taylor expansion of the $\pi^0$ TFF in Eq.~(1) should be negative
in order to agree with Eq.~(4) of the same reference.},
\begin{equation}
\begin{array}{l}
b_{\pi}=3.24(12)_{\rm{stat}}(19)_{\rm{sys}}\times 10^{-2}\ ,\\[0.5ex]
c_{\pi}=1.06(9)_{\rm{stat}}(25)_{\rm{sys}}\times 10^{-3}\ ,
\end{array}
\label{lowenergypar}
\end{equation}
where the statistical error is the result of a weighted average of several fits using different types of PAs
to the same joint set of $\pi^0$ TFF space-like data 
(see Fig.\,\ref{TFFpion}(a) for a graphical example)
and the systematic error is attributed to the model dependence of the PA method.
In this way, the values obtained for the LECs can be considered as model independent.
It is worth mentioning that the systematic errors ascribed to the LECs are quite conservative,
in the sense that they are obtained from a comparison of the constants predicted by several well-established phenomenological models
for the TFF and their counterparts extracted using various types of PAs from fits to pseudo-data sets generated by the different models.
For each LEC, the systematic error is chosen to be the largest difference among these comparisons,
making the whole approach reliable and model independent \cite{lab21}.

As one can see in Fig.\,\ref{TFFpion}(b),
the Taylor expansion in Eq.~(\ref{Taylor}) nicely describes the $\pi^{0}\to e^{+}e^{-}\gamma$ time-like experimental data \cite{lab71}.
Thus, it can be safely used for the description of the $\pi^0$ TFF in this region
within the range of available phase-space, since the first pole seems to appear, for all types of PAs considered,
inside the region of $\rho$-dominance \cite{lab21},
thus well beyond the phase-space end point.
Finally, we would like to remark that this expansion of the single-virtual $\pi^0$ TFF will be used for predicting both
$\pi^{0}\to e^{+}e^{-}\gamma$ and $\pi^{0}\to e^{+}e^{-}e^{+}e^{-}$ decays, 
the latter by means of a factorisation of the double-virtual $\pi^0$ TFF in terms of a product of the single-virtual one
(see Subsection {\ref{doublevirtual}} for details).

\subsection{$\eta\to\gamma\gamma^{*}$}\label{etaTFF}
\begin{figure*}
\centering
\includegraphics[width=0.5\textwidth]{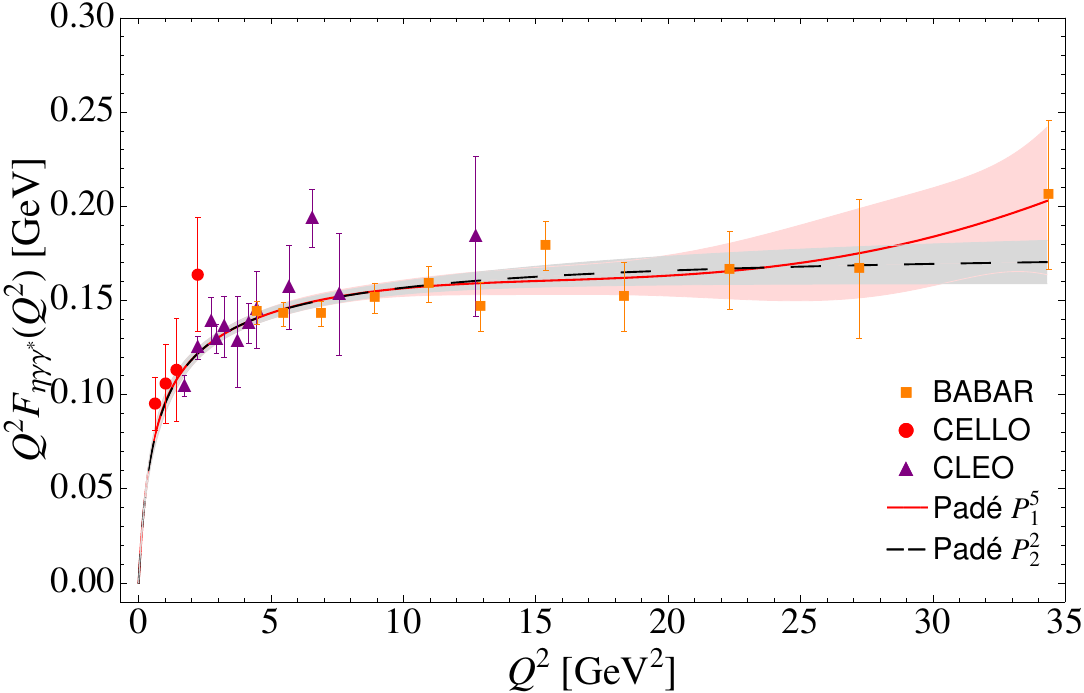}\includegraphics[width=0.5\textwidth]{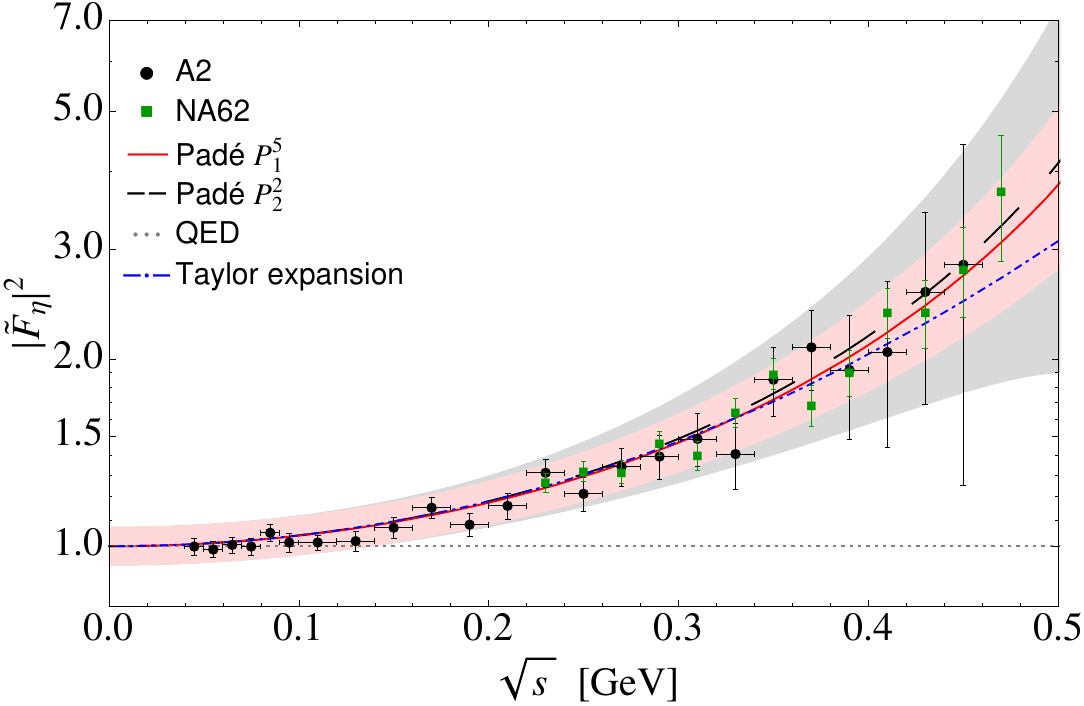}
\caption{Representation of the $\eta$ TFF as obtained from fits to the space-like experimental data using Pad\'{e} approximants \cite{lab17}
(a) and our predictions for its modulus square normalised time-like counterpart, $\widetilde F_{\eta\gamma\gamma^{\ast}}(q^{2})$,
as a function of the invariant dilepton mass, $\sqrt{s}\equiv m_{\ell\ell}$ (b).
For the $\eta$ time-like TFF, the predictions coming from the $P^5_1(q^2)$ (red solid line) and $P^2_2(q^2)$ (long-dashed black line) PAs,
and the Taylor expansion (dot-dashed blue line) are compared with the experimental data from
$\eta\to e^{+}e^{-}\gamma$ \cite{lab4} (black circles) and $\eta\to\mu^{+}\mu^{-}\gamma$ \cite{lab7} (green squares).
The one-sigma error bands associated with $P^5_1(q^2)$ (light-red) and $P^2_2(q^2)$ (light-grey) PAs are also shown. 
The QED prediction (dotted grey line) is also displayed.}
\label{TFFeta}
\end{figure*}

In order to describe the time-like region of the $\eta$ TFF from the space-like data analysis in Ref.~\cite{lab17},
we will employ two PAs, the $P^5_1(q^2)$ and the $P^2_2(q^2)$.
These are the highest-order PAs one can achieve when confronted with the joint sets of space-like experimental data
which, for the convenience of the reader, we represent in Fig.\,\ref{TFFeta}(a).
The sequence $P^L_1(q^2)$ is used when the TFF is believed to be dominated by a single resonance,
while the $P^N_N(q^2)$ one is appropriate for the case where the TFF fulfils the asymptotic behaviour\footnote{In
Ref.~\cite{lab17}, the fit to space-like data is done for $Q^2|F(Q^2)|$ and not for the TFF itself.
As a consequence, PAs satisfying the correct asymptotic limit, that is, $\lim_{Q^2\to\infty}Q^2 F(Q^2)=\mbox{const.}$,
are represented by the sequence $P^N_N(q^2)$.}.
A Taylor expansion equivalent to Eq.~(\ref{Taylor}),
with $b_{\eta}=0.60(6)_{\rm{stat}}(3)_{\rm{sys}}$ and $c_{\eta}=0.37(10)_{\rm{stat}}(7)_{\rm{sys}}$
for the slope and curvature parameters, respectively,
is better not to be used in this case because of the larger phase-space available, 
$4m_{\ell}^2\le q^2\le M_\eta^2$.
From the analysis in Ref.~\cite{lab17}, we also obtained that the fitted poles for the $P^L_1(q^2)$ sequence
are seen in the range $(0.71,0.77)$ GeV, beyond the phase-space end point, thus  again making our approach applicable 
and the predictions reliable.

Our predictions for the modulus squared of the normalised time-like $\eta$ TFF, $\widetilde F_{\eta\gamma\gamma^{\ast}}(q^{2})$,
as a function of the invariant dilepton mass, $\sqrt{s}\equiv m_{\ell\ell}$, are shown in Fig.\,\ref{TFFeta}(b),
together with the experimental data points from the A2 Collaboration for the decay $\eta\to e^+e^-\gamma$ \cite{lab4} (black circles)
and the NA60 experiment for $\eta\to\mu^+\mu^-\gamma$ \cite{lab7}\footnote{We 
thank S.~Damjanovic from the NA60 experiment for providing us with the time-like TFF data points obtained from 
$\eta\to\mu^{+}\mu^{-}\gamma$.} (green squares).
The predictions from the $P^5_1(q^2)$ (solid red  line) and $P^2_2(q^2)$ (long-dashed black line) are almost identical and in nice agreement
with the experimental data, whereas the Taylor expansion (dot-dashed blue line) is not so precise in the upper part of the spectrum. 
For this reason, we will use both PAs indiscriminately in our analysis .
The one-sigma error bands associated with $P^5_1(q^2)$ and $P^2_2(q^2)$ PAs are displayed in light-red and light-grey, respectively.
These error bands are built from the uncertainty in the coefficients of the PAs\footnote{The
coefficients of the PAs along with their errors and the correlation matrix can be obtained from the authors upon request.}
and the normalisation factor extracted from the two-photon decay width. 
These bands also include a (tiny) source attributed to the systematic uncertainty coming from the differing results,
as obtained from the element where we stop the sequence and the preceding ones.

\subsection{$\eta^{\prime}\to\gamma\gamma^{*}$}\label{etapTFF}
\begin{figure*}
\centering
\includegraphics[width=0.5\textwidth]{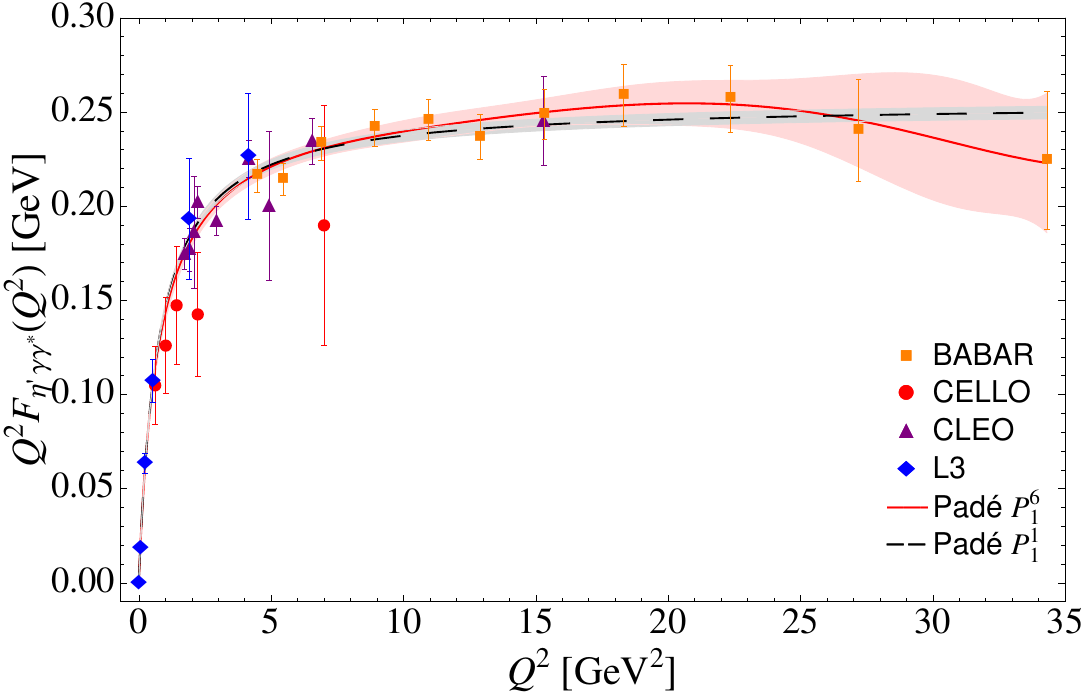}\includegraphics[width=0.5\textwidth]{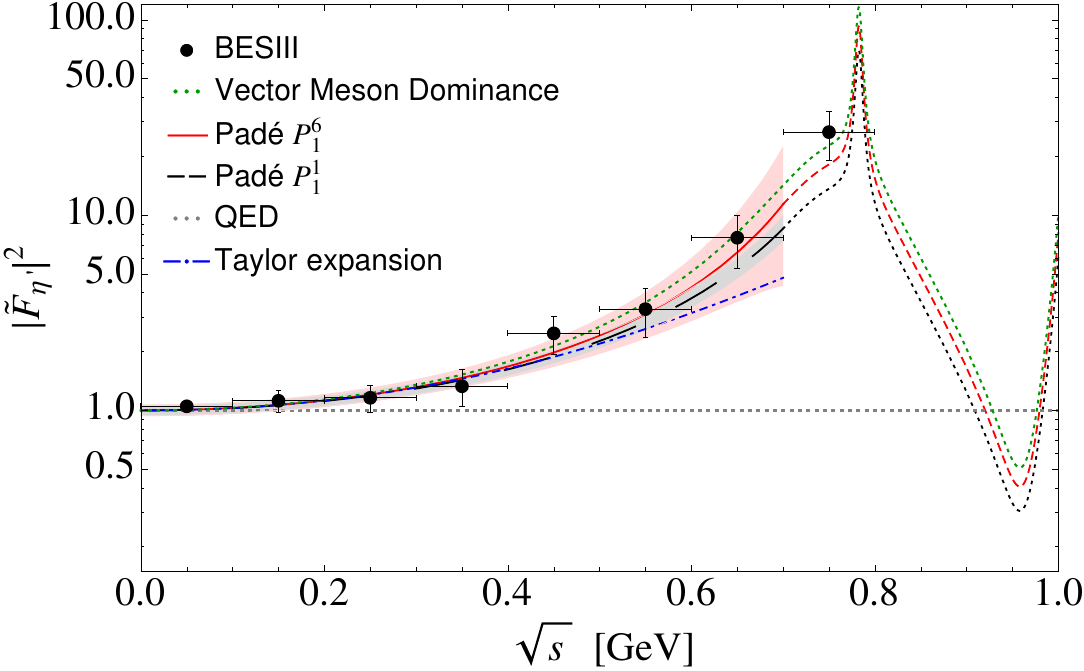}
\caption{Representation of the $\eta^{\prime}$ TFF as obtained from fits to the space-like experimental data using Pad\'{e} approximants \cite{lab17}
(a) and our predictions for its modulus squared normalised time-like counterpart, $\widetilde F_{\eta^{\prime}\gamma\gamma^{\ast}}(q^{2})$,
as a function of the invariant dilepton mass, $\sqrt{s}\equiv m_{\ell\ell}$ (b).
For the $\eta^{\prime}$ time-like TFF, the predictions up to the matching point located at $\sqrt{s}=0.70$ GeV
coming from the $P^6_1(q^2)$ (solid red line) and $P^1_1(q^2)$ (long-dashed black line) PAs,
and the Taylor expansion (dot-dashed blue line) are compared to the experimental data from $\eta^\prime\to e^{+}e^{-}\gamma$ \cite{lab8} (black circles).
From the matching point on, rescaled versions of the VMD description in Eq.~(\ref{VMD}) are used.
The one-sigma error bands associated to $P^6_1(q^2)$ (light-red) and $P^1_1(q^2)$ (light-grey) PAs are also shown. 
The QED (dotted grey line) and VMD (dotted green line) predictions are also displayed.}
\label{TFFetap}
\end{figure*}

The description of the whole time-like $\eta^{\prime}$ TFF by means of PAs is elaborate.
The available phase space, $4m_{\ell}^2\le q^2\le M_{\eta^{\prime}}^2$,
now includes  an energy region where poles associated with these PAs can emerge.
The analysis of the $\eta^{\prime}$ TFF space-like data performed in Ref.~\cite{lab17}
(see Fig.\,\ref{TFFetap}(a) for the corresponding space-like fit results)
revealed the appearance of a pole in the range $(0.83,0.86)$ GeV for the cases of a $P^L_1(q^2)$ sequence.
Consequently, we cannot employ the method of PAs for describing the time-like TFF in the entire phase-space region,
and a complementary approach must be used.
Therefore, we propose to match the description based on PAs to that given by Eq.~(\ref{VMD}) at a certain energy point\footnote{To 
proceed with the matching, we have considered an energy-dependent width for the $\rho$ resonance,
$$
\Gamma_{\rho}(q^{2})=\Gamma_{\rho}\frac{q^{2}}{M_{\rho}^{2}}\frac{\sigma^{3}(q^{2})}{\sigma^{3}(M_{\rho}^{2})}\ ,
$$
with $\sigma(q^{2})=\sqrt{1-4M^{2}_{\pi}/q^{2}}$, and a constant width for the $\omega$ and $\phi$ narrow resonances.
Input values for the masses and widths as well as for the rest of the couplings entering Eq.~(\ref{VMD}) are taken from
Ref.~\cite{lab1}.}.
Given the mass and the width of the $\rho$ meson, the first of the resonances included in the VMD description,
the region of influence due to its presence may be defined using the half-width rule as $M_\rho\pm\Gamma_\rho/2$ \cite{lab50},
thus deducing the value of the radius $\varepsilon$ mentioned earlier.
The particular energy point located at $\sqrt{s}\simeq 0.70$ GeV,
the lowest value of the former region for $M_\rho\simeq 775$ MeV and $\Gamma_\rho\simeq 150$ MeV,
fixes the optimal matching point\footnote{The
region of influence attributed to the $\omega$ and $\phi$ poles is negligible, since these are narrow resonances and
are placed far from the matching point.}.
With this value fixed, a representation valid in the whole phase-space domain is that given by the PA below the matching point
and Eq.~(\ref{VMD}) above it.
In order to match both descriptions of the form factor at the matching point we have to rescale the shape of the VMD spectrum accordingly.
In this manner, we keep track of the resonant behaviour in the upper part of the spectrum where PAs cannot be applied,
while the low-energy region is predicted in a more systematic way as compared to VMD by PAs established uniquely from space-like data.
This will allow us to integrate the whole spectrum and predict the branching ratio of the several $\eta^\prime$ Dalitz decays considered here.

Our predictions for the time-like $\eta^\prime$ TFF, together with the experimental data points from the BESIII Collaboration
for the decay $\eta^\prime\to e^+e^-\gamma$ \cite{lab8} (black circles), are displayed in Fig.\,\ref{TFFetap}(b).
The results from the $P^6_1(q^2)$ (red solid line) and $P^1_1(q^2)$ (black long-dashed line) are shown up to the matching point.
The corresponding error bands are in light-red and light-grey, respectively.
From the matching point on, our predictions are replaced by a rescaled VMD representation based on the
three lowest-lying vector resonances.
Our PA-based predictions are again in fine agreement with experiment.
A Taylor expansion with $b_{\eta^\prime}=1.30(15)_{\rm{stat}}(7)_{\rm{sys}}$ and 
$c_{\eta^\prime}=1.72(47)_{\rm{stat}}(34)_{\rm{sys}}$ \cite{lab17}
and the VMD parameterization, Eq.\,(\ref{VMD}), are also included for comparison.
It is worth mentioning that an extrapolation of the $P^6_1(q^2)$ PA beyond the matching point nicely passes through
the last experimental point.
This can be understood in the following terms.
The VMD description includes three resonances whose poles are located at different places,
while the $P^L_1(q^2)$ PAs include only one.
Making use of the single-pole approximation in Eq.~(\ref{poleFF}) and the Taylor expansion in Eq.~(\ref{Taylor})
for the case of the $\eta^\prime$,
the slope parameter is identified as $b_{\eta^\prime}=m_{\eta^\prime}^2/\Lambda^2$.
Using the $b_{\eta^\prime}$ value deduced from Eq.~(\ref{VMD}) one gets $\Lambda=M_{\rm eff}=0.822(58)$ GeV,
where the error is due to the half-width rule and can be utilised as a measure of the region of influence of the pole.
The former value is very similar to the one obtained from the pole position of the $P^6_1(q^2)$ PA, located at $\sqrt{s}=0.833$ GeV.
Therefore, the region of influence of this pole can be estimated to be in the interval $(0.77,0.89)$ GeV.
It is for this reason that the last experimental point would  also be in agreement with the $P^6_1(q^2)$ prediction \cite{lab45}.
This is not so for the $P^1_1(q^2)$ PA, thus showing that increasing the Pad\'e order allows for a better description of the data.
In any case, for the numerical analysis of the different decays involving the $\eta^\prime$, we keep both PAs for the sake of comparison.

\subsection{$\mathcal{P}\to\gamma^{\ast}\gamma^{\ast}$}\label{doublevirtual}

The double-virtual TFF, $F_{\mathcal{P}\gamma^{\ast}\gamma^{\ast}}(q_{1}^{2},q_{2}^{2})$,
depends on both photon virtualities, $q_{1}$ and $q_{2}$.
Due to Bose symmetry, it must satisfy
$F_{\mathcal{P}\gamma^{\ast}\gamma^{\ast}}(q_{1}^{2},q_{2}^{2})=F_{\mathcal{P}\gamma^{\ast}\gamma^{\ast}}(q_{2}^{2},q_{1}^{2})$.
Its normalisation is obviously the same as the single-virtual TFF,
$F_{\mathcal{P}\gamma^{\ast}\gamma^{\ast}}(0,0)=F_{\mathcal{P}\gamma\gamma^{\ast}}(0)$,
and can be extracted either from the two-photon partial width by means of Eq.~(\ref{Pgammagamma})
or from the axial anomaly.
It must also satisfy that when one of the photons is put on-shell the double-virtual TFF becomes the single-virtual one,
i.e.~$\lim_{q_i^{2}\to 0}F_{\mathcal{P}\gamma^{\ast}\gamma^{\ast}}(q_{1}^{2},q_{2}^{2})=F_{\mathcal{P}\gamma\gamma^{\ast}}(q^{2})$
for $i=1,2$.
In addition, the double-virtual TFF can fulfil the asymptotic space-like constraints,
$\lim_{Q^2\to\infty}F_{\mathcal{P}\gamma^{\ast}\gamma^{\ast}}(-Q^2,0)\propto 1/Q^2$ \cite{lab51}
and
$\lim_{Q^2\to\infty}F_{\mathcal{P}\gamma^{\ast}\gamma^{\ast}}(-Q^2,-Q^2)\propto 1/Q^2$ \cite{lab19}.

Due to the lack of experimental information in the case of double-virtual TFFs,
our initial ansatz will be to use the standard factorisation approach, which in terms of normalised form factors reads
$\widetilde F_{\mathcal{P}\gamma^{*}\gamma^{*}}(q_{1}^{2},q_{2}^{2})=
\widetilde F_{\mathcal{P}\gamma\gamma^{*}}(q_{1}^{2},0)\widetilde F_{\mathcal{P}\gamma\gamma^{*}}(0,q_{2}^{2})$
\cite{lab52,lab53,lab54}.
This double-virtual TFF description may or may not satisfy the high-energy constraints above.
As an example, the PA $P_{1}^{0}(q^{2})=a_{0}/(1-a_{1}q^{2})$, corresponding to the single-pole approximation in Eq.~(\ref{poleFF})
motivated by VMD,
would induce a $1/q^{4}$ term in the double-virtual TFF, which violates the last of the asymptotic constraints mentioned before
(OPE prediction)
\cite{lab19,lab55,lab56,lab57}.
For this reason, we also use for our study the lowest order bivariate approximant
\begin{equation}
P_{1}^{0}(q_{1}^{2},q_{2}^{2})=
\frac{a_{0,0}}{1-\frac{b_{1,0}}{M_{\mathcal{P}}^{2}}(q_{1}^{2}+q_{2}^{2})+\frac{b_{1,1}}{M_{\mathcal{P}}^{4}}q_{1}^{2}q_{2}^{2}}\ ,
\label{Chisholm}
\end{equation}
which consists of a generalisation of the univariate PAs named Chisholm approximants (CAs) \cite{lab46}.
The analysis of the $\pi^{0}\to e^{+}e^{-}$ decay is a recent example that illustrates the application of these CAs \cite{lab58}
(see also P.~Masjuan's contribution in Ref.~\cite{lab59}).
In Eq.~(\ref{Chisholm}), 
$a_{0,0}$ is identified as the normalisation $F_{\mathcal{P}\gamma^\ast\gamma^\ast}(0,0)$ and then fixed from Eq.~(\ref{Pgammagamma}), 
$b_{1,0}$ is the slope of the single-virtual TFF obtained in Refs.~\cite{lab17,lab21},
that is, $b_{\pi}$ from Eq.~(\ref{lowenergypar}) for the pion and 
$b_{\eta^{(\prime)}}$ from Eq.~(5) in Ref.~\cite{lab17} for the $\eta$ and $\eta^{\prime}$, respectively,
and $b_{1,1}$ corresponds to the double-virtual slope which may be extracted in the future as soon as experimental data
for the double-virtual TFFs become available.
For the numerical analysis, we consider, as a conservative estimate, 
varying $b_{1,1}$ from the value respecting the OPE prediction, $b_{1,1}=0$, to $b_{1,1}=2b_{1,0}^{2}$,
far from the factorisation result $b_{1,1}=b_{1,0}^{2}$.
In this manner, we can test the sensitivity of our predictions to the double-virtual slope.
We also encourage experimental groups to perform double-virtual TFF measurements in order to fix this parameter.
In this work, we employ both descriptions indiscriminately, the factorisation ansatz and the bivariate approximant in Eq.~(\ref{Chisholm}).
See also Ref.~\cite{lab60} for a recent approach to the double-virtual TFF of the $\eta$ meson
based on the standard factorisation approach.

\section{Single Dalitz decays}\label{Single}

The generic amplitude for the single Dalitz decays involves the corresponding single-virtual TFF, as described in Section {\ref{FormFactor}},
and reads
\begin{eqnarray}
\mathcal{A}(\mathcal{P}\to\ell^{+}\ell^{-}\gamma)&=&
-ie^{2}F_{\mathcal{P}\gamma\gamma^{*}}(q^2)
\varepsilon^{\alpha\beta\mu\nu}k_{\alpha}q_{\beta}\epsilon_{\mu}^{*}(k)\\
&&\times\frac{-ig_{\nu\rho}}{q^2}\bar{u}(p_{\ell^{-}})(-ie)\gamma^{\rho}v(p_{\ell^{+}})\ .\nonumber
\end{eqnarray}
The corresponding differential decay rate is given by
\begin{eqnarray}
\frac{d\Gamma_{\mathcal{P}\to\ell^{+}\ell^{-}\gamma}}{d\sqrt{s}\,\Gamma^{\rm exp}_{\mathcal{P}\to\gamma\gamma}}&=&
\frac{4\alpha}{3\pi\sqrt{s}}\sqrt{1-\frac{4m_{\ell}^{2}}{s}}\left(1+\frac{2m_{\ell}^{2}}{s}\right)\\
&&\times\left(1-\frac{s}{M_{\mathcal{P}}^{3}}\right)^{3}|\widetilde{F}_{\mathcal{P}\gamma\gamma^{*}}(s)|^{2}\ ,\nonumber
\end{eqnarray}
where $s\equiv q^2=(p_{+}+p_{-})^2\equiv M_{\ell^{+}\ell^{-}}^2$ is the dilepton invariant mass
and the TFF appears to be normalised in order to avoid misunderstandings due to the different conventions for
$F_{\mathcal{P}\gamma\gamma^{*}}(0)$ existing in the literature.
For the numerical computations, we have employed the PrimEx Coll.~resulting value
$\Gamma_{\pi^{0}\to\gamma\gamma}=7.82(14)(17)$ eV \cite{lab61}
and the PDG fitted values $\Gamma_{\eta\to\gamma\gamma}=0.516(18)$ keV and
$\Gamma_{\eta^{\prime}\to\gamma\gamma}=4.35(14)$ keV \cite{lab1}.

\subsection{$\pi^{0}\to e^{+}e^{-}\gamma$}

The decay process $\pi^{0}\to e^{+}e^{-}\gamma$ was suggested for the first time by Dalitz in 1951~\cite{lab62}.
The first branching ratio prediction arose from a pure QED radiative corrections calculation neglecting
the momentum dependence of the TFF \cite{lab63}.
These radiative corrections have been revisited recently in Ref.~\cite{lab64}.
When our description for the differential decay rate distribution in terms of the dilepton invariant mass
is compared to the QED pointlike prediction, the agreement is found to be almost perfect.
The reason is that the main contribution to this rate comes from the very low-energy part of the mass spectrum,
where the effect of the TFF is negligible\footnote{We
arbitrarily define the very low-, low- and high-energy part of the spectrum as the part contributing 50\%, 30\% and 20\%, respectively,
to the integrated decay rate.}.
Tiny differences appear solely in the high-energy part of the spectrum.
Our result for the integrated branching ratio is 1.174(12)\%,
as compared to the QED pointlike value 1.172\%.
Our prediction is in excellent agreement with the PDG reported value $(1.174\pm 0.035)\%$~\cite{lab1}.
The main source of error we have quoted arises from the uncertainty associated with the low energy constants in Eq.~(\ref{lowenergypar}).
The error related to the measured two-photon decay width \cite{lab61} is also included.
Our result is in accord with several theoretical predictions existing in the
literature~\cite{lab70,lab33,lab35,lab36}
and the QED estimates of Refs.~\cite{lab33,lab65,lab66}.
Needless to say, the dimuon mode in the final state is kinematically forbidden in this case.

\subsection{$\eta\to\ell^{+}\ell^{-}\gamma$ $(\ell=e,\mu)$}

\begin{table*}
\caption{Branching ratio predictions for $\eta\to\ell^{+}\ell^{-}\gamma$ $(\ell=e, \mu)$ and their comparisons with
experimental measurements and previous theoretical calculations.}
\label{tabledalitzeta}
\centering
\begin{tabular}{ccc}
\hline\noalign{\smallskip}
Source & $\mathcal{BR}(\eta\to e^{+}e^{-}\gamma)\times 10^{3}$ & $\mathcal{BR}(\eta\to\mu^{+}\mu^{-}\gamma)\times 10^{4}$\\
\noalign{\smallskip}\hline\noalign{\smallskip}
QED & $6.38$ & $2.17$\\[0.5ex]
This work $(P_{1}^{5})$ & $6.60^{+0.50}_{-0.47}$ & $3.25^{+0.40}_{-0.36}$\\[0.5ex]
This work $(P_{2}^{2})$ & $6.61^{+0.53}_{-0.49}$ & $3.30^{+0.65}_{-0.56}$\\
\noalign{\smallskip}\hline\noalign{\smallskip}
PDG \cite{lab1} & $6.9\pm 0.4$ & $3.1\pm 0.4$\\[0.5ex]
H.~Berghauser {\it et al.} \cite{lab3} & $6.6\pm 0.4_{\rm{stat}}\pm 0.4_{\rm{syst}}$ & \\[0.5ex]
WASA-at-COSY Coll.~\cite{lab6} & $6.72\pm 0.07_{\rm{stat}}\pm 0.31_{\rm{syst}}$ & \\
\noalign{\smallskip}\hline\noalign{\smallskip}
QED \cite{lab66} & $6.38$ & $2.18$\\[0.5ex]
FF 2,3 \cite{lab34} & $6.57$ & $3.05$\\[0.5ex]
FF 4 \cite{lab34} & $6.53$ & $2.87$\\[0.5ex]
LFQM \cite{lab35} & $6.95$ & $2.94$\\[0.5ex]
Hidden gauge \cite{lab36} & $6.57\pm 0.03$ & $3.05\pm 0.04$\\[0.5ex]
Modified VMD \cite{lab36} & $6.55\pm 0.03$ & $2.97\pm 0.05$\\
\noalign{\smallskip}\hline
\end{tabular}
\end{table*}

Due to the larger mass of the $\eta$ as compared to the $\pi^0$, the dimuon mode in the final state is also accessible in this case.
For that reason, these decays are more challenging for testing the momentum dependence of the TFF,
and higher deviations from the QED pointlike estimates are expected.
This is precisely what our predictions reflect in Fig.~\ref{Dalitzeta}, where the dilepton invariant mass distributions of
$\eta\to e^{+}e^{-}\gamma$ (solid blue curve) and $\eta\to\mu^{+}\mu^{-}\gamma$ (solid black curve)
are compared to the QED predictions (dotted and dashed grey curves, respectively).
For the sake of comparison, we have employed a single-pole $P_{1}^{5}(q^{2})$ PA in our prediction.
The diagonal (asymptotically well-behaved) PA $P_{2}^{2}(q^{2})$ produces a very similar description,
as expected from the small differences between the two PA versions of the TFF in the whole kinematical range,
as discussed in Section {\ref{etaTFF}}.
With one version or the other, the impact of the TFF is much bigger, in absolute terms, in the dimuon case,
the reason being again a peaked distribution in the very-low energy region of the dielectron spectrum,
which gives the most important contribution to the branching ratio, where the effect of the TFF is negligible.
Notice that the high-energy parts of the spectra overlap, since the only difference between them is the dilepton production threshold.
Once the distributions are integrated,
we see from Table \ref{tabledalitzeta} that the branching ratio of the dimuon mode has been increased by $50\%$ 
as compared to its prediction in QED, while the effect in the dielectron case is not sizeable
(compatible with the QED calculation, within errors).
The asymmetric errors shown in Table \ref{tabledalitzeta} arise from the error bands displayed in Fig.~\ref{TFFeta}.
Our predictions are in agreement with present experimental measurements.
They also agree with previous theoretical calculations,
such as the QED predictions in Ref.~\cite{lab66},
the results in Ref.~\cite{lab34} using different form factors (FF) based on constraints from QCD and on experiments,
the values in Ref.~\cite{lab35} from the light-front quark model (LFQM),
and those obtained in Ref.~\cite{lab36} within the hidden gauge and the modified VMD models.

\begin{center}
\includegraphics[width=0.45\textwidth]{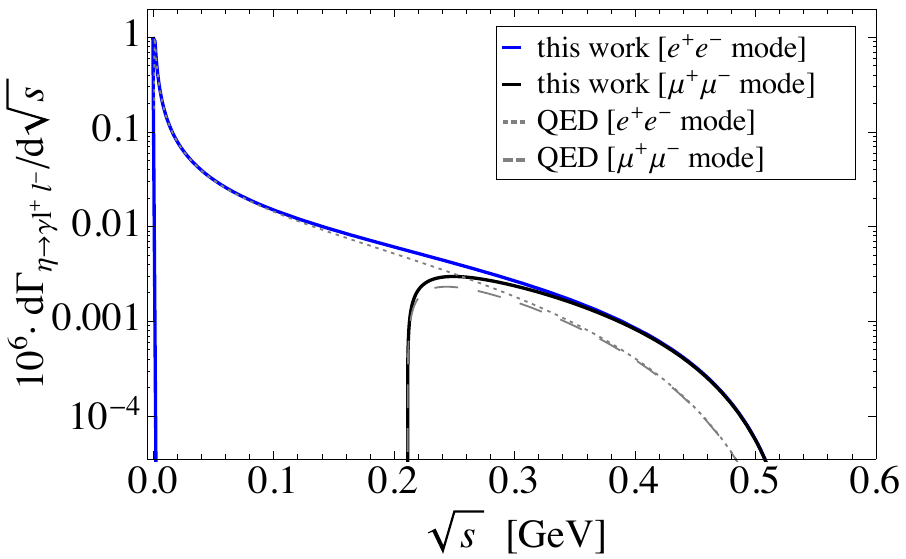}
\figcaption{\label{Dalitzeta}Decay rate distribution for $\eta\to e^{+}e^{-}\gamma$ (solid blue  curve) and $\eta\to\mu^{+}\mu^{-}\gamma$ (solid black  curve).
The corresponding QED estimates are also displayed (dotted and long-dashed grey curves, respectively).}
\end{center}

\subsection{$\eta^{\prime}\to\ell^{+}\ell^{-}\gamma$ $(\ell=e,\mu)$}

\begin{table*}
\caption{Branching ratio predictions for $\eta^{\prime}\to\ell^{+}\ell^{-}\gamma$ $(\ell=e, \mu)$ and their comparisons with
experimental measurements and previous theoretical calculations.}
\label{tabledalitzetap}
\centering
\begin{tabular}{ccc}
\hline\noalign{\smallskip}
Source & $\mathcal{BR}(\eta^{\prime}\to e^{+}e^{-}\gamma)\times 10^{4}$ & 
$\mathcal{BR}(\eta^{\prime}\to\mu^{+}\mu^{-}\gamma)\times 10^{4}$\\
\noalign{\smallskip}\hline\noalign{\smallskip}
QED & $3.94$ & $0.38$\\[0.5ex]
This work $(P_{1}^{6})$ & $4.42^{+0.39}_{-0.35}$ & $0.81^{+0.16}_{-0.13}$\\[0.5ex]
This work $(P_{1}^{1})$ & $4.35^{+0.29}_{-0.27}$ & $0.74\pm 0.06$\\
\noalign{\smallskip}\hline\noalign{\smallskip}
PDG \cite{lab1,lab9} & & $1.08\pm 0.27$\\[0.5ex]
BESIII Coll.~\cite{lab8} & $4.69\pm 0.20_{\rm{stat}}\pm 0.23_{\rm{syst}}$ & \\
\noalign{\smallskip}\hline\noalign{\smallskip}
Hidden gauge \cite{lab36} & $4.62\pm 0.17$ & $0.98\pm 0.05$\\[0.5ex]
Modified VMD \cite{lab36} & $4.53\pm 0.17$ & $0.90\pm 0.05$\\
\noalign{\smallskip}\hline
\end{tabular}
\end{table*}

Due to the even larger mass of the $\eta^\prime$, the phase space is now enlarged by around 400 MeV
and the associated TFF could be explored up to virtualities of the order of 1 GeV$^2$ if decay distributions of these processes
were available.
Unfortunately, this is not the case and predictions for these distributions must be provided.
Our proposal for the $\eta^\prime$ TFF is the one discussed in Section {\ref{etapTFF}},
which consists in using the PA results obtained from space-like data up to the matching point located at $\sqrt{s}=0.70$ GeV,
then supplemented by a VMD description including the $\rho$, $\omega$ and $\phi$ vector resonances beyond that point.
In Fig.~\ref{Dalitzetap}, the dilepton invariant mass distributions of
$\eta^\prime\to e^{+}e^{-}\gamma$ (solid blue curve) and $\eta^\prime\to\mu^{+}\mu^{-}\gamma$ (solid black curve)
are compared to the corresponding QED predictions (dotted and dashed grey curves, respectively).
For such a comparison, we have employed for our prediction a single-pole $P_{1}^{6}(q^{2})$ PA.
The description achieved from the diagonal PA $P_{1}^{1}(q^{2})$ is quite similar, as anticipated from the observation of Fig.~\ref{TFFetap}.
For the dielectron case, the decay distribution evidences again a marked peak at low energy which,
despite the contribution coming from the resonance region, dominates the branching ratio,
as occurred in $\pi^{0}(\eta)\to e^{+}e^{-}\gamma$.
However, the effect of the TFF on the $\eta^{\prime}\to\mu^{+}\mu^{-}\gamma$ decay distribution is larger than in 
$\eta\to\mu^{+}\mu^{-}\gamma$, increasing the branching ratio by a factor of 2.
This is because of both a larger phase space and the effect of passing through a $q^{2}$ region
where resonances can be produced on-shell.
Interestingly, the contribution of the $\rho$ resonance bends the distribution, while the inclusion of the $\omega$ resonance accounts for
the sharp peak around 0.8 GeV.
Numerical predictions for the branching ratios are presented in Table~\ref{tabledalitzetap},
where the errors shown come from the error bands linked to the TFF.
Our predictions are in accordance with the theoretical calculations in Ref.~\cite{lab36},
while they are slightly below the recent experimental measurement of $\eta^{\prime}\to e^{+}e^{-}\gamma$ \cite{lab8}
and the old measurement of $\eta^{\prime}\to\mu^{+}\mu^{-}\gamma$~\cite{lab9},
though in agreement within errors in both cases.

\begin{center}
\includegraphics[width=0.45\textwidth]{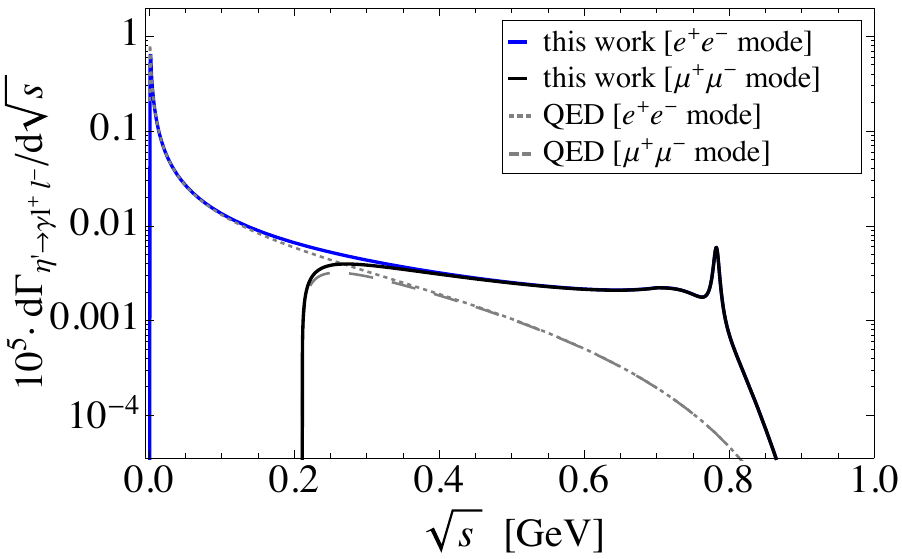}
\figcaption{\label{Dalitzetap}Decay rate distribution for $\eta^{\prime}\to e^{+}e^{-}\gamma$ (solid blue  curve) and
$\eta^{\prime}\to\mu^{+}\mu^{-}\gamma$ (solid black  curve).
The corresponding QED estimates are also displayed (dotted and long-dashed grey curves, respectively).}
\end{center}

\begin{center}
\includegraphics[width=0.2\textwidth]{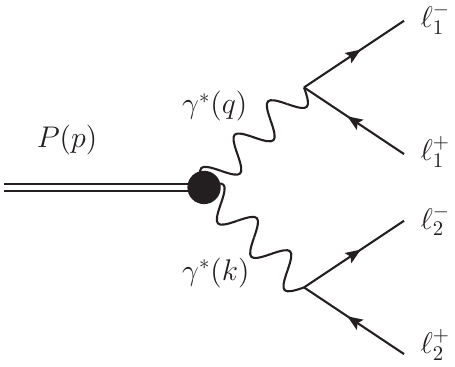}
\quad
\includegraphics[width=0.2\textwidth]{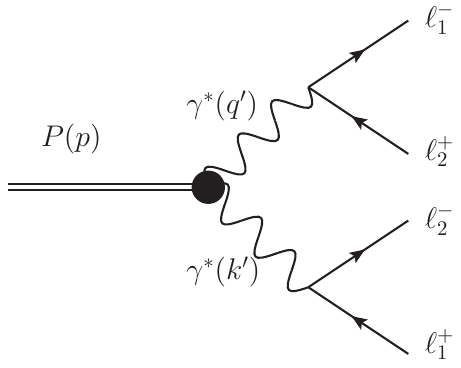}
\figcaption{\label{DoubleDalitz}Direct (a) and exchange (b) diagrams associated with double Dalitz decays.}
\end{center}

\begin{table*}
\caption{Branching ratio predictions for $\pi^{0}\to e^{+}e^{-}e^{+}e^{-}$ and their comparisons with experimental measurements.}
\label{pi04e}
\centering
\begin{tabular}{lllcc}
\hline\noalign{\smallskip}
\multirow{2}{*}{Source} & 
\multicolumn{2}{c}{\multirow{2}{*}{Double-virtual TFF}} & 
\multicolumn{2}{l}{$\mathcal{BR}(\pi^{0}\to e^{+}e^{-}e^{+}e^{-})\times 10^5$}\\[0.5ex]
& \multicolumn{2}{c}{} & direct+exchange & interference\\
\noalign{\smallskip}\hline\noalign{\smallskip}
\multirow{6}{*}{This work} & 
\multirow{3}{*}{Chisholm approximants} & $b_{1,1}=0$ & 3.42937(9) & -0.03599\\[0.5ex]
& & $b_{1,1}=b_{1,0}^2$ & 3.42937(9) & -0.03599\\[0.5ex]
& & $b_{1,1}=2b_{1,0}^2$ & 3.42936(9) & -0.03599\\[0.5ex]
& Factorisation ansatz & eq.~(\ref{Taylor}) & 3.42945(9) & -0.03599\\[0.5ex]
& QED & & 3.41607 & -0.03484\\
\noalign{\smallskip}\hline\noalign{\smallskip}
\multicolumn{3}{l}{\multirow{2}{*}{Experimental measurements}} &
\multicolumn{2}{l}{3.38(16) \cite{lab1}}\\[0.5ex]
\multicolumn{3}{l}{} &
\multicolumn{2}{l}{3.46(19) \cite{lab10}}\\
\noalign{\smallskip}\hline  
\end{tabular}
\end{table*}

\section{Double Dalitz decays}
\label{Double}

The double Dalitz decays involve the double-virtual TFF, as described in Section~{\ref{FormFactor}}.
They involve four leptons in the final state, which makes the phase-space integration much more tedious.
In the case of having two pairs of non-identical leptons, that is $\eta^{(\prime)}\to e^{+}e^{-}\mu^{+}\mu^{-}$,
the required diagram is shown in Fig.~\ref{DoubleDalitz}(a) and the decay amplitude reads
\begin{eqnarray}
\label{amplitude2e2mu}
\lefteqn{\mathcal{A}_{\eta^{(\prime)}\to e^{+}e^{-}\mu^{+}\mu^{-}}=
-i\frac{e^{4}}{q^{2}k^{2}}F_{\eta^{(\prime)}\gamma^{*}\gamma^{*}}(q^{2},k^{2})}\\
&\times&\varepsilon^{\alpha\beta\mu\nu}k_{\alpha}q_{\beta}
\bar{u}(q_{e^{-}})\gamma_{\mu}v(q_{e^{+}})\bar{u}(q_{\mu^{-}})\gamma_{\nu}v(q_{\mu^{+}})\ ,\nonumber
\end{eqnarray}
where $q^2=(q_{e^{+}}+q_{e^{-}})^2\equiv M_{e^{+}e^{-}}^2$
and $k^2=(q_{\mu^{+}}+q_{\mu^{-}})^2\equiv M_{\mu^{+}\mu^{-}}^2$
are the dielectron and dimuon invariant masses, respectively.
The corresponding differential decay rate can be reduced to\footnote{See, for instance,
Ref.~\cite{lab67} for reducing four-body final-state distributions into two invariant masses.}
\begin{eqnarray}
\label{distribution}
\lefteqn{\displaystyle
\frac{d^{2}\Gamma_{\eta^{(\prime)}\to e^{+}e^{-}\mu^{+}\mu^{-}}}
{dM^{2}_{e^{+}e^{-}}M^{2}_{\mu^{+}\mu^{-}}\,\Gamma^{\rm exp}_{\eta^{(\prime)}\to\gamma\gamma}}=
\mathcal{S}\frac{8\alpha^{2}}{9\pi^{2}m_{\eta^{(\prime)}}^{6}}\frac{\displaystyle 1}{k^{2}q^{2}}}\nonumber\\ 
&\times&\sqrt{1-\frac{4m_{e}^{2}}{k^{2}}}\sqrt{1-\frac{4m_{\mu}^{2}}{q^{2}}}
\left(1+\frac{2m^{2}_{e}}{k^{2}}\right)\left(1+\frac{2m^{2}_{\mu}}{q^{2}}\right)\nonumber\\
&\times&
\left\{\frac{1}{4}\left[m_{\eta^{(\prime)}}^{2}-(k^{2}+q^{2})\right]^2-k^{2}q^{2}\right\}^{3/2}
|\widetilde{F}(k^{2},q^{2})|^{2}\ ,
\end{eqnarray}
where, in this case, $\mathcal{S}=2$, in agreement with the expression given in Ref.~\cite{lab36}.
In the case of having two pairs of identical leptons in the final state,
that is $\mathcal{P}\to e^{+}e^{-}e^{+}e^{-}$ or $\eta^{(\prime)}\to\mu^{+}\mu^{-}\mu^{+}\mu^{-}$,
one must instead consider both the direct (a) and exchange (b) diagrams in Fig.~\ref{DoubleDalitz}.
In this case, the total amplitude of the process then reads
\begin{equation}
\mathcal{A}=\mathcal{A}_{\rm{dir}}-\mathcal{A}_{\rm{exch}}\ ,
\label{direx}
\end{equation}
where the minus sign appears due to the exchange of the two identical leptons.
Squaring the former amplitude, one arrives at
\begin{equation}
|\mathcal{A}|^{2}=|\mathcal{A}_{\rm{dir}}|^{2}+|\mathcal{A}_{\rm{exch}}|^{2}-2\Re(\mathcal{A}_{\rm{dir}}\mathcal{A}^{*}_{\rm{exch}})\ ,
\label{interference}
\end{equation}
where not only the contributions from both the direct and exchange diagrams appear, but also that of the interference term.
The contributions to the total partial decay width from the first and second terms of Eq.~(\ref{interference}) are obviously the same,
that is, $\Gamma_{\rm{dir}}=\Gamma_{\rm{exch}}$.
In this way, the decay width $\Gamma_{\rm{dir+exch}}$ is obtained to be that of Eq.~(\ref{distribution}), now with $\mathcal{S}=1$,
in accordance with Ref.~\cite{lab41},
once the factor $\frac{1}{2!2!}$ accounting for the two pairs of identical particles in the final state and the sum over the two contributions
have been taken into account.
Regarding the interference term, a detailed expression in terms of five invariant masses, which requires a Monte Carlo (MC) simulation
to be integrated, is relegated to the appendix.

\begin{center}
\includegraphics[width=0.45\textwidth]{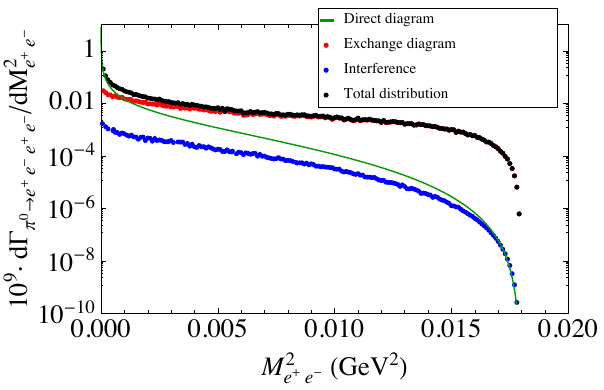}
\figcaption{\label{DoubleDalitzpi0}Different contributions to the $\pi^{0}\to e^{+}e^{-}e^{+}e^{-}$ decay rate distribution
as a function of one dielectron invariant mass of the direct diagram:
direct diagram (solid green  curve), exchange diagram (dotted red  curve), interference term (dotted blue  curve), 
and total distribution (dotted black  curve).}
\end{center}

\begin{table*}
\caption{Branching ratio predictions for $\eta\to e^{+}e^{-}\mu^{+}\mu^{-}$ and their comparison with the current experimental upper bound.}
\label{eta2e2mu}
\centering
\begin{tabular}{lllc}
\hline\noalign{\smallskip}
Source &
\multicolumn{2}{c}{Double-virtual TFF} & $\mathcal{BR}(\eta\to e^{+}e^{-}\mu^{+}\mu^{-})\times 10^6$\\
\noalign{\smallskip}\hline\noalign{\smallskip}
\multirow{7}{*}{This work} & \multirow{3}{*}{Chisholm approximants} & $b_{1,1}=0$ & 2.39(12)\\[0.5ex]
& & $b_{1,1}=b_{1,0}^2$ & 2.39(12)\\[0.5ex]
& & $b_{1,1}=2b_{1,0}^2$ & 2.38(12)\\[0.5ex]
& \multirow{2}{*}{Factorisation ansatz} & $P_{1}^{5}$ & $2.35^{+0.47}_{-0.40}$\\[0.5ex]
& & $P_{2}^{2}$ & $2.39^{+0.66}_{-0.53}$\\[0.5ex]
& QED & & 1.57\\
\noalign{\smallskip}\hline\noalign{\smallskip}
\multicolumn{3}{l}{Experimental measurement} & $<1.6\times 10^{-4}$ (90\% CL) \cite{lab5}\\
\noalign{\smallskip}\hline  
\end{tabular}
\end{table*} 

\subsection{$\pi^{0}\to e^{+}e^{-}e^{+}e^{-}$}

The only possible double Dalitz decay of the neutral pion is $\pi^{0}\to e^{+}e^{-}e^{+}e^{-}$. Other possibilities are not kinematically allowed.
In view of the results from $\pi^{0}\to e^{+}e^{-}\gamma$, one may expect that the overall effect of the TFF will be again small.
In Fig.~\ref{DoubleDalitzpi0}, we show our results for the different contributions to the decay rate distribution
as a function of the invariant mass of one dielectron pair of the direct diagram.
Concretely, we display the curve corresponding to the direct diagram (green solid line),
the curve of the contribution of the exchange diagram expressed in terms of the former dielectron invariant mass
of the direct diagram\footnote{The 
curve of the exchange diagram expressed in its own variables would look equal to the solid green  line of Fig.~\ref{DoubleDalitzpi0}.
In this work, we have opted to show, in just one figure, all the contributions as a function of one dielectron invariant mass
of the direct diagram.
In this convention, the exchange diagram as expressed in Fig.~\ref{DoubleDalitzpi0} has also required a MC integration.}
(dotted red  line), the interference term (dotted blue  line) and finally the total distribution (dotted black  line).
We want to note that the contribution from both direct and exchange diagrams integrates in the same way and that
the interference is small and destructive.
Our $\mathcal{BR}$ predictions are shown in Table~\ref{pi04e}, from which we corroborate that the effect of the TFF is small
because the main contribution to the $\mathcal{BR}$ proceeds from the very low-momentum transferred region where a peak emerges,
as already occurred in $\pi^{0}\to e^{+}e^{-}\gamma$.
Our results are in good agreement with current experimental measurements.
The source of the associated error comes from the uncertainty on the low-energy parameters in Eq.~(\ref{lowenergypar}).
Notice that the sensitivity of this decay to the variations of the double-virtual slope parameter, $b_{1,1}$, is at the fifth decimal place.
In this sense, the high level of accuracy demanded to infer its value is unthinkable at the current level of precision.
It is also interesting to compare with other authors' results.
We are in good agreement with the results given in Refs.~\cite{lab36,lab41,lab53}
for the direct and exchange contributions, while the result of Ref.~\cite{lab35} is about $5\%$ lower than our predictions.
Regarding the interference term, we have a perfect agreement with Refs.~\cite{lab36,lab53}
and a value about $30\%$ higher than
Ref.~\cite{lab41}, while Ref.~\cite{lab35} did not consider this term.
Comparing with previous QED estimates, we agree with Refs.~\cite{lab65,lab66} for the direct and exchange contributions.
For the interference term the former did not consider it and the latter gave a result 5 times larger than ours.

\subsection{$\eta\to\ell^{+}\ell^{-}\ell^{+}\ell^{-}$ $(\ell=e,\mu)$}

The double Dalitz decays of the $\eta$,
$\eta\to e^{+}e^{-}e^{+}e^{-}$, $\eta\to \mu^{+}\mu^{-}\mu^{+}\mu^{-}$ and $\eta\to e^{+}e^{-}\mu^{+}\mu^{-}$ processes are now kinematically allowed.
Let us first analyse the latter, for simplicity.
In this case, the two dilepton pairs are different and consequently there is no interference phenomenon.
Hence, the distribution rate is just given by Eq.~(\ref{distribution}) and shown in Fig.~\ref{DoubleDalitzeta2e2mu} in two different manners,
one expressed in terms of the dielectron invariant mass and the other in the dimuon variable (blue and red solid lines respectively),
where, of course, both curves integrate the same.
Our predictions are shown in Table~\ref{eta2e2mu},
where the source of the associated errors comes from the error bands associated with the TFF
for the case of the factorisation approach, and from the uncertainty on the single-virtual slope for the description employing CAs.
From the experimental side, we respect the current upper limit, while from the theory side,
because of the appearance of a dimuon pair in the final state,
the effect of the TFF increases the $\mathcal{BR}$ about $50\%$, for the same arguments as explained for $\eta\to\ell^{+}\ell^{-}\gamma$.
This decay, though much more sensitive than $\pi^{0}\to e^{+}e^{-}e^{+}e^{-}$ to the double-virtual slope, $b_{1,1}$,
would require accurate measurements as well as demanding a very precise description of the TFF,
in order to diminish its associated error, for deducing $b_{1,1}$, far from the present situation.
Comparing with other authors, we agree with the predictions of Ref.~\cite{lab36},
while we have found discrepancies with the value $5.83\times 10^{-7}$ of Ref.~\cite{lab35},
with the prediction $2\times 10^{-7}$ of Ref.~\cite{lab34} and with the estimate $7.84\times 10^{-7}$ of Ref.~\cite{lab66}.
In the latter case, the reason seems to be a typographical fault of a factor of 2 missing, as pointed out
in both Refs.~\cite{lab36,lab53}.
In such a case, it would reproduce the QED result of Table~\ref{eta2e2mu} as it should be,
because they did not consider the momentum dependence of the TFF.

\begin{center}
\includegraphics[width=0.45\textwidth]{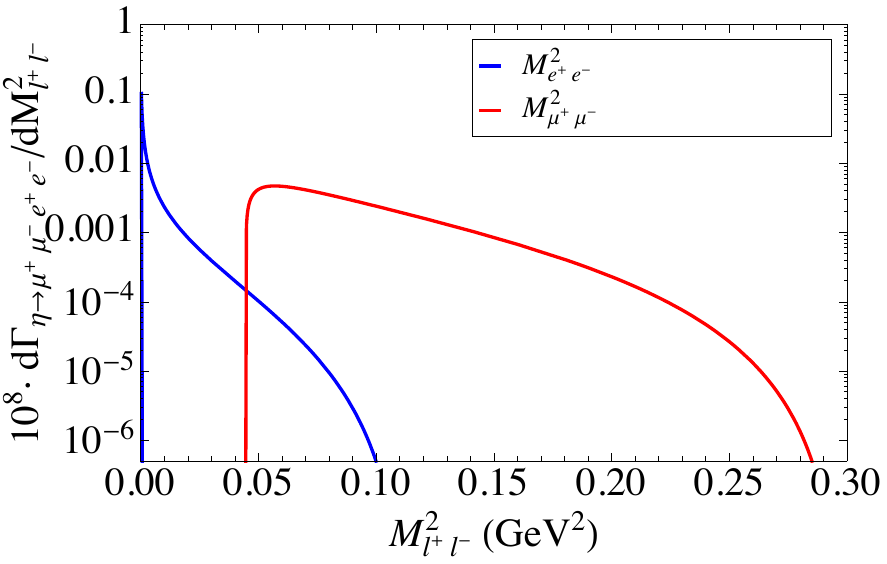}
\figcaption{\label{DoubleDalitzeta2e2mu}Decay rate distribution for $\eta\to e^{+}e^{-}\mu^{+}\mu^{-}$
as a function of dielectron (blue curve) and dimuon (red curve) invariant mass.}
\end{center}

\begin{table*}
\caption{Branching ratio predictions for $\eta\to e^{+}e^{-}e^{+}e^{-}$ and $\eta\to\mu^{+}\mu^{-}\mu^{+}\mu^{-}$
and their comparisons with experimental measurements.}
\label{eta4l}
\centering
\begin{tabular}{lllcccc}
\hline\noalign{\smallskip}
\multirow{2}{*}{Source} & 
\multicolumn{2}{c}{\multirow{2}{*}{Double-virtual TFF}} & 
\multicolumn{2}{l}{$\mathcal{BR}(\eta\to e^{+}e^{-}e^{+}e^{-})\times 10^5$} & 
\multicolumn{2}{l}{$\mathcal{BR}(\eta\to\mu^{+}\mu^{-}\mu^{+}\mu^{-})\times 10^9$}\\[0.5ex]
& \multicolumn{2}{c}{} & dir+exch & interf & dir+exch & interf\\
\noalign{\smallskip}\hline\noalign{\smallskip}
\multirow{7}{*}{This work} &
\multirow{3}{*}{Chisholm approximants} & $b_{1,1}=0$ & 2.74(3) &-0.02 & 4.47(33) & -0.32\\[0.5ex]
& & $b_{1,1}=b_{1,0}^2$ & 2.73(3) & -0.03 & 4.31(31) & -0.32\\[0.5ex]
& & $b_{1,1}=2b_{1,0}^2$ & 2.73(3) & -0.03 & 4.15(30) & -0.32\\[0.5ex]
& \multirow{2}{*}{Factorisation ansatz} &
$P_{1}^{5}$ & $2.72^{+0.43}_{-0.37}$ & -0.03 & $4.23^{+0.83}_{-0.70}$ & -0.43\\[0.5ex]
& & $P_{2}^{2}$ & $2.73^{+0.45}_{-0.39}$ & -0.03 & $4.30^{+1.12}_{-0.91}$ & -0.47\\[0.5ex]
& QED & & 2.56 & -0.02 & 2.59 & -0.19\\
\noalign{\smallskip}\hline\noalign{\smallskip}
\multicolumn{3}{l}{\multirow{2}{*}{Experimental measurements}} &
\multicolumn{2}{l}{$3.2(9)_{\rm{stat}}(5)_{\rm{sys}}$ \cite{lab6}} &
\multicolumn{2}{l}{\multirow{2}{*}{$<3.6\times 10^{-4}$ (90\% CL) \cite{lab5}}}\\[0.5ex] 
\multicolumn{3}{l}{} & \multicolumn{2}{l}{$2.4(2)_{\rm{stat}}(1)_{\rm{sys}}$ \cite{lab11}} &
\multicolumn{2}{l}{}\\
\noalign{\smallskip}\hline  
\end{tabular}
\end{table*}

\begin{figure*}
\centering
\includegraphics[width=0.45\textwidth]{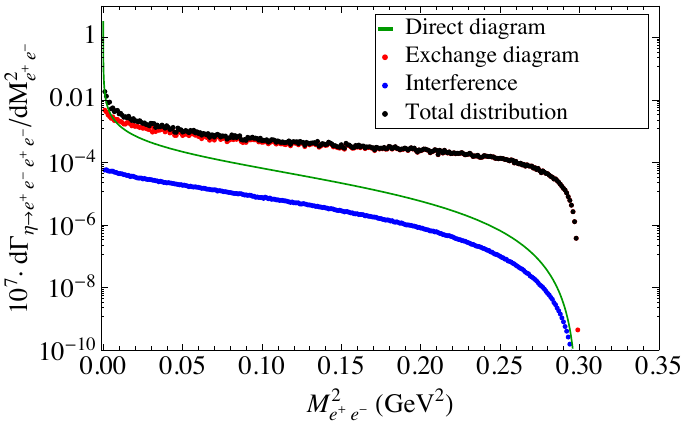}
\quad
\includegraphics[width=0.45\textwidth]{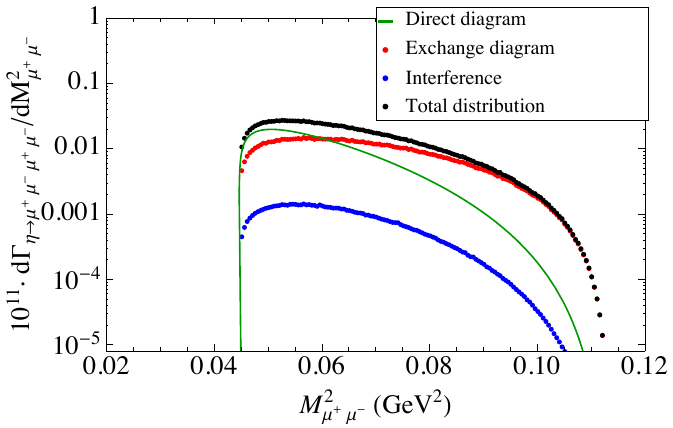}
\caption{Different contributions to the $\eta\to e^{+}e^{-}e^{+}e^{-}$ (a) and $\eta\to \mu^{+}\mu^{-}\mu^{+}\mu^{-}$ (b)
decay rate distributions as a function of one dielectron invariant mass of the direct diagram:
direct diagram (solid green  curve), exchange diagram (dotted red  curve), interference term (dotted blue  curve), 
and total distribution (dotted black  curve).}
\label{spectra4leptons}
\end{figure*}

The decays involving two identical dilepton pairs in the final state, $\eta\to e^{+}e^{-}e^{+}e^{-}$ and $\eta\to \mu^{+}\mu^{-}\mu^{+}\mu^{-}$,
require us to consider Eq.~(\ref{interference}).
Their distributions are given in Fig.~\ref{spectra4leptons} ((a) and (b) respectively)
as a function of one dilepton invariant mass of the direct diagram.
We explicitly show the contribution from the direct diagram (solid green  curve),
the curve of the exchange diagram expressed in terms of the former dielectron (dimuon) invariant mass
of the direct diagram (dotted red  curve),
the interference term (dotted blue  curve) and the total decay rate distribution (dotted black  line).
The integrated $\mathcal{BR}$ results are shown in Table~\ref{eta4l},
where the error comes again from the error bands of the TFF description as given in Fig.~\ref{TFFeta}, for the factorisation approach,
and from the uncertainty associated with the slope, $b_{\eta}$, for the description using CAs. 
Comparing with present experimental status,
our prediction for $\eta\to e^{+}e^{-}e^{+}e^{-}$ is compatible at less than $1\sigma$ with the KLOE measurement \cite{lab11},
as well as with the recent measurement value of the WASA@COSY collaboration~\cite{lab6},
while our estimate for $\eta\to\mu^{+}\mu^{-}\mu^{+}\mu^{-}$ respects the current upper bound of Ref.~\cite{lab5}.
We have found the same trend as in $\eta\to\ell^{+}\ell^{-}\gamma$, that is,
while the overall effect of the TFF on the electronic mode is small,
increasing the $\mathcal{BR}$ of $\eta\to e^{+}e^{-}e^{+}e^{-}$ by $6\%$ with respect to the QED estimate,
the impact on the muonic channel, $\eta\to\mu^{+}\mu^{-}\mu^{+}\mu^{-}$,
becomes important, increasing the $\mathcal{BR}$ by a factor ranging  (1.6--1.7) with respect to the QED calculation.
As a consequence, the sizeable sensitivity of $\eta\to\mu^{+}\mu^{-}\mu^{+}\mu^{-}$ to the TFF of double virtuality makes it
a good candidate to improve our knowledge of it.
Interestingly, a precise experimental measurement of this mode at the percent level of precision leaves us in a position
to estimate the value of $b_{1,1}$.
For that purpose, it is also required to diminish the associated uncertainty with the TFF.
Here enters the ability of the Pad\'{e} method we use for accommodating new experimental data as soon as released
by experimental groups.
However, the same exercise for $\eta\to e^{+}e^{-}e^{+}e^{-}$ would demand accurate measurements at the per mille level
to unveil this quantity, far from the present situation.
Our predictions are in good agreement with the results of Ref.~\cite{lab36} for the electronic mode,
while (10--15)\% over the muonic prediction.
Comparing with Ref.~\cite{lab35} (which did not considered the interference term) we are  $10\%$ over for the electronic case,
while his result for $\eta\to\mu^{+}\mu^{-}\mu^{+}\mu^{-}$ is $60\%$ smaller.
We are also in accordance with the estimate of Ref.~\cite{lab34} for $\eta\to e^{+}e^{-}e^{+}e^{-}$.
Regarding the pure QED calculation of Ref.~\cite{lab66},
we are in perfect agreement for the electronic channel while tiny differences are found in the muonic decay,
probably caused by the updated values of our input values.

\subsection{$\eta^{\prime}\to\ell^{+}\ell^{-}\ell^{+}\ell^{-}$ $(\ell=e,\mu)$}

Regarding the double Dalitz decays of the $\eta^{\prime}$,
we have the same three possible final states as for the $\eta$.
However, in this case we have only adopted the factorisation approach ansatz for describing the double-virtual TFF of the $\eta^{\prime}$.
The reason is because the use of Chisholm approximants, which may respect the appropriate asymptotic behaviour $q^{-2}$,
would only apply at low energies, concretely up to the matching point where PAs are applicable,
while beyond, we are somehow forced to employ the factorisation approximation, through a VMD description, which induces a $q^{-4}$ term.
So, there is no gain respecting the high-energy behaviour in the low-energy region if we violate it at high energies.
We compute first $\eta^{\prime}\to e^{+}e^{-}\mu^{+}\mu^{-}$, again through Eq.~(\ref{distribution}).
Noticeably, it follows the same trend as $\eta\to e^{+}e^{-}\mu^{+}\mu^{-}$,
with the difference that this case is sensitive to the resonance region, as can be read off from Fig.~\ref{decayetap2e2mu}.
Once more, the low-momentum region basically dominates the distribution when working with the electronic variable (solid blue  curve),
while it is a smooth falling function of the dimuonic momentum with a small bump and a sharp peak accounting for the effect of the
$\rho$ and the $\omega$, respectively (solid red  curve).
Both curves integrate to the same $\mathcal{BR}$.
Our predictions are presented in Table~\ref{etap2e2mu} without, in this case, any experimental reference to compare with.
The effect of the TFF increases by a factor of about 2 the $\mathcal{BR}$ respect to the QED estimate,
which is much notorious than in $\eta\to e^{+}e^{-}\mu^{+}\mu^{-}$.

\begin{center}
\includegraphics[width=0.45\textwidth]{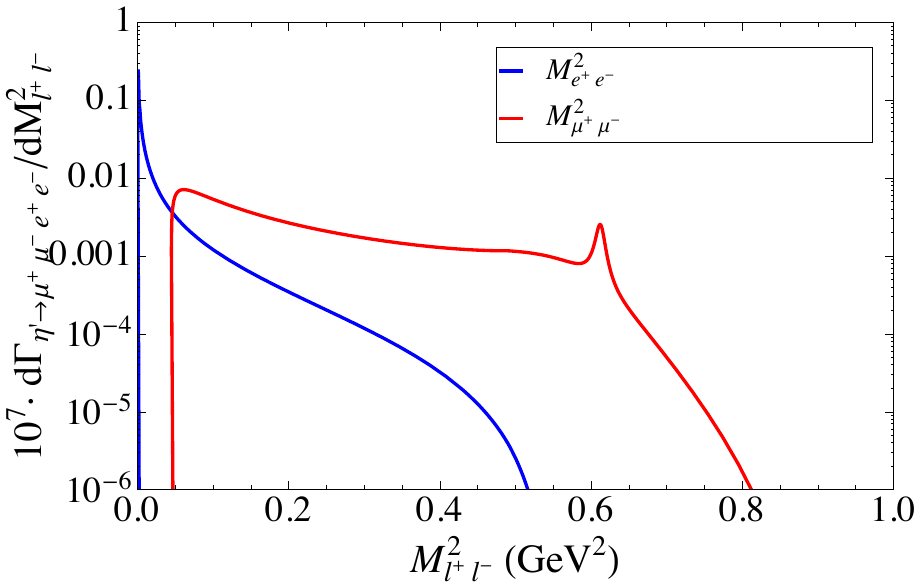}
\figcaption{\label{decayetap2e2mu}Decay rate distribution for $\eta^{\prime}\to e^{+}e^{-}\mu^{+}\mu^{-}$
as a function of dielectron (blue curve) and dimuon (red curve) invariant mass.}
\end{center}

\begin{center}
\tabcaption{Branching ratio predictions for $\eta^{\prime}\to e^{+}e^{-}\mu^{+}\mu^{-}$.}
\label{etap2e2mu}
\begin{tabular*}{80mm}{c@{\extracolsep{\fill}}ccc}
\hline\noalign{\smallskip}
Source & $\mathcal{BR}(\eta^{\prime}\to e^{+}e^{-}\mu^{+}\mu^{-})\times 10^{7}$\\
\noalign{\smallskip}\hline\noalign{\smallskip}
This work $(P_{1}^{6})$ & $6.80^{+1.39}_{-1.17}$\\[0.5ex]
This work $(P_{1}^{1})$ & $6.25^{+0.83}_{-0.72}$\\[0.5ex]
QED & $3.21$\\
\noalign{\smallskip}\hline\noalign{\smallskip}
Exp.~measurements & not seen\\
\noalign{\smallskip}\hline
\end{tabular*}
\end{center}

\begin{figure*}
\centering
\includegraphics[width=0.45\textwidth]{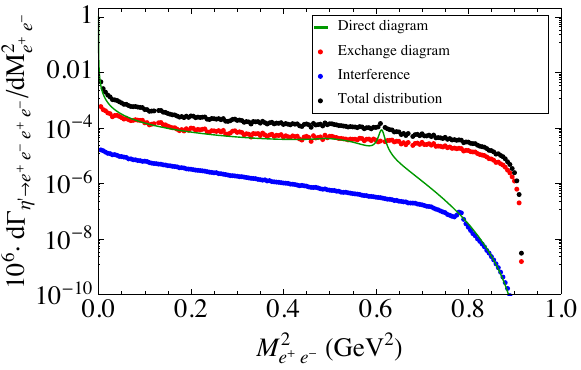}
\quad
\includegraphics[width=0.45\textwidth]{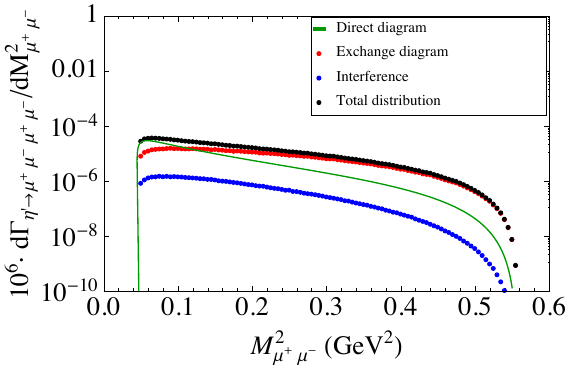}
\caption{Different contributions to the $\eta^{\prime}\to e^{+}e^{-}e^{+}e^{-}$ (a) and $\eta^{\prime}\to \mu^{+}\mu^{-}\mu^{+}\mu^{-}$ (b)
decay rate distributions as a function of one dielectron invariant mass of the direct diagram:
direct diagram (solid green  curve), exchange diagram (dotted red  curve), interference term (dotted blue  curve), 
and total distribution (dotted black  curve).}
\label{decayetap4leptons}
\end{figure*}

The decay spectra for $\eta^{\prime}\to e^{+}e^{-}e^{+}e^{-}$ and $\eta^{\prime}\to\mu^{+}\mu^{-}\mu^{+}\mu^{-}$
shown in Fig.~\ref{decayetap4leptons} ((a) and (b) respectively)
have been computed by taking Eq.~(\ref{interference}) into account.
We have represented the contributions of the direct diagram (solid green  curve),
the exchange diagram expressed in terms of the variable of the direct diagram (dotted red  curve),
the interference term (dotted blue  curve) and lastly the total distribution (dotted black  line).
One interesting feature concerning phase space is that the electronic mode (Fig.~\ref{decayetap4leptons}(a)) is clearly sensitive
to the intermediate vector resonances,
while the muonic (Fig.~\ref{decayetap4leptons}(b)) is basically not.
Our predictions are presented in Table~\ref{etap4ltable}, and also reflect the tendency that the effect of the TFF is sizeable
and larger than for the case of the $\eta$.
In particular, the $\mathcal{BR}$ of $\eta^{\prime}\to e^{+}e^{-}e^{+}e^{-}$ and $\eta^{\prime}\to\mu^{+}\mu^{-}\mu^{+}\mu^{-}$
have increased by $20\%$ and by a factor of 2, respectively.
On the experimental side, we have no observations to compare with,
while on the theory side we have only found the predictions given in Ref.~\cite{lab36}
with which we are in good agreement for the cases with two identical dilepton pairs in the final state,
while we are slightly below for $\eta^{\prime}\to\mu^{+}\mu^{-}e^{+}e^{-}$. 

\begin{table*}
\caption{Branching ratio predictions for $\eta^{\prime}\to e^{+}e^{-}e^{+}e^{-}$ and $\eta^{\prime}\to\mu^{+}\mu^{-}\mu^{+}\mu^{-}$.}
\label{etap4ltable}
\centering
\begin{tabular}{lllcccc}
\hline\noalign{\smallskip}
\multirow{2}{*}{Source} & 
\multicolumn{2}{c}{\multirow{2}{*}{Double-virtual TFF}} & 
\multicolumn{2}{l}{$\mathcal{BR}(\eta^{\prime}\to e^{+}e^{-}e^{+}e^{-})\times 10^6$} & 
\multicolumn{2}{l}{$\mathcal{BR}(\eta^{\prime}\to\mu^{+}\mu^{-}\mu^{+}\mu^{-})\times 10^8$}\\[0.5ex]
& \multicolumn{2}{c}{} & dir+exch & interf & dir+exch & interf\\
\noalign{\smallskip}\hline\noalign{\smallskip}
\multirow{3}{*}{This work} 
& \multirow{2}{*}{Factorisation ansatz} &
$P_{1}^{6}$ & $2.15^{+0.35}_{-0.29}$ & $-0.03$ & $2.19^{+0.23}_{-0.19}$ & $-0.44$\\[0.5ex]
& & $P_{1}^{1}$ & $2.09^{+0.28}_{-0.24}$ & $-0.01$ & $2.06^{+0.17}_{-0.15}$ & $-0.41$\\[0.5ex]
& QED & & $1.75$ & $-0.01$ & $0.98$ & $-0.11$\\
\noalign{\smallskip}\hline\noalign{\smallskip}
\multicolumn{3}{l}{Experimental measurements} &
\multicolumn{2}{c}{not seen} &
\multicolumn{2}{c}{not seen}\\
\noalign{\smallskip}\hline  
\end{tabular}
\end{table*}

\begin{table*}
\caption{Central final branching ratio predictions as a combined weighted average of the results presented.
Errors are symmetrised. $n_{\sigma}$ stands for the number of standard deviations the measured results are from our predictions.}
\label{centralresults}
\centering
\begin{tabular}{lccc}
\hline\noalign{\smallskip}
Decay&This work&Experimental value \cite{lab1}&$n_{\sigma}$\\
\noalign{\smallskip}\hline\noalign{\smallskip}
$\pi^{0}\to e^{+}e^{-}\gamma$&$1.169(1)\%$&$1.174(35)\%$&0.15\\[0.5ex]
$\eta\to e^{+}e^{-}\gamma$&$6.61(50)\times 10^{-3}$&$6.90(40)\times 10^{-3}$&0.45\\[0.5ex]
$\eta\to\mu^{+}\mu^{-}\gamma$&$3.26(46)\times 10^{-4}$&$3.1(4)\times 10^{-4}$&0.26\\[0.5ex]
$\eta^{\prime}\to e^{+}e^{-}\gamma$&$4.38(32)\times 10^{-4}$&$4.69(20)(23)\times 10^{-4}$&0.70\\[0.5ex]
$\eta^{\prime}\to\mu^{+}\mu^{-}\gamma$&$0.75(6)\times 10^{-4}$&$1.08(27)\times 10^{-4}$&1.19\\[0.5ex]
$\pi^{0}\to e^{+}e^{-}e^{+}e^{-}$&$3.36689(5)\times 10^{-5}$&$3.34(16)\times 10^{-5}$&0.17\\[0.5ex]
$\eta\to e^{+}e^{-}e^{+}e^{-}$&$2.71(2)\times 10^{-5}$&$2.4(2)(1)\times 10^{-5}$&1.38\\[0.5ex]
$\eta\to\mu^{+}\mu^{-}\mu^{+}\mu^{-}$&$3.98(15)\times 10^{-9}$&$<3.6\times 10^{-4}$\\[0.5ex]
$\eta\to e^{+}e^{-}\mu^{+}\mu^{-}$&$2.39(7)\times 10^{-6}$&$<1.6\times 10^{-4}$\\[0.5ex]
$\eta^{\prime}\to e^{+}e^{-}e^{+}e^{-}$&$2.10(45)\times 10^{-6}$&not seen\\[0.5ex]
$\eta^{\prime}\to\mu^{+}\mu^{-}\mu^{+}\mu^{-}$&$1.69(36)\times 10^{-8}$&not seen\\[0.5ex]
$\eta^{\prime}\to e^{+}e^{-}\mu^{+}\mu^{-}$&$6.39(91)\times 10^{-7}$&not seen\\
\noalign{\smallskip}\hline
\end{tabular}
\end{table*}

\section{Conclusions}
\label{Conclusions}

The single and double Dalitz decays $\mathcal{P}\to\ell^{+}\ell^{-}\gamma$ and $\mathcal{P}\to\ell^{+}\ell^{-}\ell^{+}\ell^{-}$
($\mathcal{P}=\pi^{0}, \eta, \eta^{\prime}$; $\ell=e$ or $\mu$) have been analysed by means of a data-driven
model-independent description of the transition $\mathcal{P}\to\gamma^{(*)}\gamma^{*}$.
We have benefited from (our) previous findings on the space-like single-virtual TFF $\gamma\gamma^{*}\to\mathcal{P}$
obtained through the use of Pad\'{e} approximants to represent these transitions in the time-like energy region where they are applicable.
We have shown that this extrapolation from the space-like to the time-like is supported by current experimental data
$\eta$ and $\eta^\prime$ TFFs obtained from $\eta^{(\prime)}\to e^{+}e^{-}\gamma$ and $\eta\to\mu^{+}\mu^{-}\gamma$ decays.
This nice behaviour proves that these TFFs are well approximated by meromorphic functions.
Regarding the TFF of double virtuality, besides the standard factorisation approach, we have motivated the use of bivariate approximants, 
which would satisfy the high-energy constraints and whose coefficients may be determined as soon as experimental data become available.
From the phenomenological point of view, 
we have found that the invariant mass distributions involving electrons in the final state show strong peaks
in the very low-momentum transfer region,
which mainly dominate the contribution to the branching ratios, hence suppressing the effect of the TFFs.
However, distributions involving muons in the final state are much more homogeneously distributed and
clearly manifest the neat effect of the TFF, which, in particular, is enhanced for the $\eta^{\prime}$ decays
due to phase-space considerations.
Our final branching ratio predictions are summarised in Table \ref{centralresults}, 
where a combined weighted average\footnote{For
the combined weighted average we follow the standard procedure for unconstrained averaging described in Ref.~\cite{lab1},
that is to say,
$\overline{\mathcal{BR}}\pm \delta\overline{\mathcal{BR}}=
\left(\sum_{i}\mathcal{BR}_{i}/(\delta\mathcal{BR}_{i})^{2}/\sum_{i}1/(\delta\mathcal{BR}_{i})^{2}\right)\pm
\left(\sum_{i}1/(\delta\mathcal{BR}_{i})^{2}\right)^{-1/2}$,
where $\mathcal{BR}_{i}$ and $\delta\mathcal{BR}_{i}$ are the values and errors displayed in the different tables
for the same particular branching ratio.}
of the results given in the different tables has been considered and the uncertainties symmetrised.
The values of $n_{\sigma}$ shown in the table give the number of standard deviations
the experimental measurements are from our predictions.
All these predictions are seen to be in accordance with present experimental measurements;
only $\eta^{\prime}\to\mu^{+}\mu^{-}\gamma$ and $\eta\to e^{+}e^{-}e^{+}e^{-}$ appear slightly in tension.
It is worth stating that our results for the $\eta$ and $\eta^\prime$ decays are independent of
$\eta$-$\eta^\prime$ mixing effects, the reason being that all the PAs used to fit the space-like TFFs are fixed at zero momentum transfer
by the corresponding experimental two-photon decay widths and not by the axial-anomaly predictions
in terms of the mixing parameters.
Similarly, when diagonal PAs were used, a constant behaviour for $Q^2F(Q^2)$ was imposed at $Q^2\to\infty$,
as predicted by perturbative QCD,
but without fixing the associated constants in terms of these mixing parameters.
To finish, we would like once more to encourage experimental groups to measure these TFFs.

\acknowledgments{We thank very much
Pere Masjuan, Santi Peris, Pablo Roig and Pablo S\'anchez Puertas for fruitful discussions on this topic. 
We are grateful to the NA60 Collaboration, in particular to Sanja Damjanovic, 
for providing us with the transition form factor data obtained from $\eta\to\mu^{+}\mu^{-}\gamma$.
S.~Gonz\`alez-Sol\'is would also like to thank the Institut f\"{u}r Kernphysik of the 
Johannes Gutenberg-Universit\"{a}t in Mainz for hospitality.
This work was supported in part by the FPI scholarship BES-2012-055371 (S.G-S),
the Secretaria d'Universitats i Recerca del Departament d'Economia i Coneixement de la Generalitat de Catalunya
under grant 2014 SGR 1450,
the Ministerio de Ciencia e Innovaci\'on under grant FPA2011-25948,
the Ministerio de Econom\'{i}a y Competitividad under grants CICYT-FEDER-FPA 2014-55613-P and SEV-2012-0234,
the Spanish Consolider-Ingenio 2010 Program CPAN (CSD2007-00042),
and the European Commission under program FP7-INFRASTRUCTURES-2011-1 (Grant Agreement N. 283286).
S.G-S also received support from the CAS President's International Fellowship Initiative for Young International Scientists (Grant No. 2017PM0031), by the Sino-German Collaborative Research Center \textquotedblleft Symmetries and the Emergence of Structure in QCD\textquotedblright\,(NSFC Grant No.\,11621131001, DFG Grant No.\,TRR110), and by NSFC (Grant No.\,11647601).}

\end{multicols}

\newpage

\begin{multicols}{2}

\subsection*{Appendix: Interference term}\label{interferenceterm}
\begin{small}

\noindent{\bf Four-body decay width in invariant variables}

The partial decay width of a particle $P$ of mass $M_P$ decaying into four particles $p_{1}p_{2}p_{3}p_{4}$ 
reads \cite{lab1}
\begin{eqnarray}
\label{decaywidth}
\lefteqn{\Gamma(P\to p_{1}p_{2}p_{3}p_{4})=
\int d\Phi(p_{P};q_{1},q_{2},q_{3},q_{4})}\\
&\times&\frac{(2\pi)^{4}}{2M_{P}}
\overline{\big|\mathcal{M}(P\to p_{1}p_{2}p_{3}p_{4})\big|^{2}}\ ,\nonumber
\end{eqnarray}
where $d\Phi(p_{P};q_{1},q_{2},q_{3},q_{4})$ is the four-body phase-space element given by
\begin{eqnarray}
\lefteqn{d\Phi(p_{P};q_{1},q_{2},q_{3},q_{4})=
\delta^4\left(p_P-\sum\limits_{i=1}^{4}q_{i}\right)}\nonumber\\
&\times&\prod_{i=1}^{4}\frac{d^{3}\mathbf{q}_{i}}{(2\pi)^{3}2E_{i}}\ .
\end{eqnarray}
Following Refs.~\cite{lab68,lab69}, 
the phase space is expressed in terms of independent invariant masses (instead of using three-momenta and angles) as
\begin{eqnarray}
\lefteqn{d\Phi(p_{P};q_{1},q_{2},q_{3},q_{4})=
\frac{1}{8\pi^{10}M_{P}^{2}}}\\
&\times&\left(-B\right)^{-1/2}dM_{12}^{2}dM_{34}^{2}dM_{14}^{2}dM_{124}^{2}dM_{134}^{2}\ ,\nonumber
\end{eqnarray}
where $M_{ij}^2=(q_{i}+q_{j})^{2}$ and $M_{ijk}^2=(q_{i}+q_{j}+q_{k})^{2}$.
In the cases that concern us, that is, the interference term for $\pi^{0}\to e^{+}e^{-}e^{+}e^{-}$ and
$\eta^{(\prime)}\to\ell^{+}\ell^{-}\ell^{+}\ell^{-}$ $(\ell=e,\mu)$, $B$ reads
\begin{eqnarray}
\lefteqn{B=m_{\ell}^{8}+\Big[M_{124}^{2}\left(M_{134}^{2}-M_{34}^{2}\right)+M_{14}^{2}\left(-M_{P}^{2}+M_{34}^{2}\right)\Big]^{2}}
\nonumber\\
&&+M_{12}^{2}\Big\{M_{12}^{2}\left(M_{134}^{2}-M_{14}^{2}\right)^{2}\nonumber\\
&&-2\left(M_{134}^{2}-M_{14}^{2}\right)\left(M_{124}^{2}M_{134}^{2}-M_{P}^{2}M_{14}^{2}\right)\nonumber\\
&&+2\Big[\left(-2M_{P}^{2}+M_{134}^{2}-M_{14}^{2}\right)M_{14}^{2}\nonumber\\
&&+M_{124}^{2}\left(M_{134}^{2}+M_{14}^{2}\right)\Big]M_{34}^{2}\Big\}\nonumber\\
&&+m_{\ell}^{4}\Big\{M_{12}^{4}+M_{124}^{4}+4M_{124}^{2}M_{134}^{2}+M_{134}^{4}\nonumber\\
&&-2M_{P}^{2}\Big[2\left(M_{124}^{2}+M_{134}^{2}\right)+M_{14}^{2}\Big]\nonumber\\
&&+2\left(4M_{124}^{2}+M_{134}^{2}+M_{14}^{2}\right)M_{34}^{2}+M_{34}^{4}\nonumber\\
&&+2M_{12}^{2}\left(M_{124}^{2}+4M_{134}^{2}+M_{14}^{2}+M_{34}^{2}\right)\Big\}\nonumber\\
&&+m_{\ell}^{6}\Big[4M_{P}^{2}-2\left(3M_{12}^{2}+M_{124}^{2}+M_{134}^{2}+3M_{34}^{2}\right)\Big]\nonumber\\
&&-2m_{\ell}^{2}\Big\{M_{12}^{4}\left(M_{134}^{2}+M_{14}^{2}\right)\nonumber\\
&&+\left(2M_{P}^{2}-M_{124}^{2}-M_{134}^{2}\right)\left(-M_{124}^{2}M_{134}^{2}+M_{P}^{2}M_{14}^{2}\right)\nonumber\\
&&+\Big[M_{124}^{4}+\left(-3M_{P}^{2}+M_{124}^{2}+M_{134}^{2}\right)M_{14}^{2}\Big]M_{34}^{2}\nonumber\\
&&+\left(M_{124}^{2}+M_{14}^{2}\right)M_{34}^{4}\nonumber\\
&&+M_{12}^{2}\Big[M_{34}^{4}-3M_{P}^{2}M_{14}^{2}+M_{124}^{2}M_{14}^{2}+M_{134}^{2}M_{14}^{2}\nonumber\\
&&+\left(M_{124}^{2}+M_{134}^{2}-2M_{14}^{2}\right)M_{34}^{2}\Big]\Big\}\ ,
\end{eqnarray}
where $m_{\ell}$ is the lepton mass and the boundary of the physical allowed region fulfils $B=0$.
Reference~\cite{lab68} points out that the choice of variables $M_{12}^{2}, M_{34}^{2}, M_{14}^{2}, M_{124}^{2}$, 
and $M_{134}^{2}$ is convenient to facilitate the finding of the integration limits of $B$, since it
only depends quadratically on each of the variables.
Other choices can lead to quartics.\\[1ex]

\noindent{\bf Integration limits}

In order to find the physical region of one variable, for instance $M_{14}^{2}$, one must solve $B=0$, obtaining
\begin{eqnarray}
\lefteqn{M_{14}^{2\pm}=
\frac{1}{\lambda(M_{12}^{2},M_{34}^{2},M_{P}^{2})}}\\
&\times&\left[-b\pm2\sqrt{G(M_{124}^{2},M_{34}^{2},M_{12}^{2},m_{\ell}^{2},M_{P}^{2},m_{\ell}^{2})}\right.\nonumber\\
&&\left.\times\sqrt{G(M_{134}^{2},M_{34}^{2},M_{12}^{2},m_{\ell}^{2},M_{P}^{2},m_{\ell}^{2})}\right]\ ,\nonumber
\end{eqnarray}
where $\lambda(a,b,c)=a^{2}+b^{2}+c^{2}-2ab-2ac-2bc$ is the basic two-particle kinematical function,
the K\'allen function, 
$b$ is given by
\begin{eqnarray}
\lefteqn{b=(M_{34}^{2}-M_{12}^{2})\Big[M_{34}^{2}(M_{124}^{2}+m_{\ell}^{2})}\\
&-&(M_{124}^{2}-m_{\ell}^{2})(M_{134}^{2}-m_{\ell}^{2})-2m_{\ell}^{4}\Big]\nonumber\\
&-&M_{12}^{2}\Big[(M_{124}^{2}-m_{\ell}^{2})(M_{134}^{2}-m_{\ell}^{2})\nonumber\\
&+&M_{34}^{2}(M_{124}^{2}+M_{134}^{2}+2m_{\ell}^{2}-2M_{P}^{2})\nonumber\\
&+&(M_{134}^{2}+3m_{\ell}^{2})M_{P}^{2}\Big]
+(M_{12}^{2})^{2}(M_{134}^{2}+m_{\ell}^{2})\ ,\nonumber
\end{eqnarray}
and
\begin{eqnarray}
\lefteqn{G(x,y,z,u,v,w)=u^2 z-u v w}\\ 
&+&u v x-u v z+u w y-u w z-u x y-u x z-u y z\nonumber\\
&+&u z^2+v^2 w+v w^2 -v w x-v w y-v w z-v x y\nonumber\\
&+&v y z-w x y+w x z+x^2 y+x y^2-x y z\ ,\nonumber
\end{eqnarray}
is the basic four-particle kinematic function.
As argued in Ref.~\cite{lab68}, the integration limits of the remaining variables, 
$M_{12}^{2}, M_{34}^{2}, M_{124}^{2}$, and $M_{134}^{2}$ are obtained after solving
\begin{eqnarray}
\label{G0}
G(M_{124}^{2},M_{34}^{2},M_{12}^{2},m_{\ell}^{2},M_{P}^{2},m_{\ell}^{2})&=&0\ ,\\
G(M_{134}^{2},M_{34}^{2},M_{12}^{2},m_{\ell}^{2},M_{P}^{2},m_{\ell}^{2})&=&0\ ,\nonumber
\end{eqnarray}
while the dilepton invariant masses $M_{12}^{2}$ and $M_{34}^{2}$ range from threshold
$4m_{\ell}^{2}$ to $(M_{P}-2m_{\ell})^2$ and $4m_{\ell}^{2}$ to $(M_{P}-M_{12})^2$, respectively.\\[1ex]

\noindent{\bf Matrix element of the interference term}

The last term in Eq.~(\ref{interference}) reads
\begin{eqnarray}
\label{amplitude}
\lefteqn{\mathcal{A}_{1}\mathcal{A}_{2}=
\frac{\displaystyle e^{8}|F(q^{2},k^{2})||F(q^{\prime2},k^{\prime2})|}{\displaystyle q^{2}k^{2}q^{\prime2}k^{\prime2}}
\varepsilon^{\mu\nu\alpha\beta}\varepsilon^{\mu^{\prime}\nu^{\prime}\alpha^{\prime}\beta^{\prime}}}\\
&\times&(q_{1}+q_{2})_{\mu}(q_{3}+q_{4})_{\nu}(q_{1}+q_{4})_{\mu^{\prime}}(q_{2}+q_{3})_{\nu^{\prime}}\nonumber\\[0.5ex]
&\times&\mathrm{Tr}[(\slashed{q_{1}}+m_{\ell})\gamma_{\alpha}(\slashed{q_{2}}-m_{\ell})\gamma_{\beta^{\prime}}
(\slashed{q_{3}}+m_{\ell})\gamma_{\beta}(\slashed{q_{4}}-m_{\ell})\gamma_{\alpha^{\prime}}]\ .\nonumber
\end{eqnarray}

The trace and the corresponding contractions with both the product of Levi-Civita tensors and the different 
diphoton four momenta in Eq.~(\ref{amplitude}) have been computed with FormCalc.
To give a result in the chosen variables, one needs to perform some replacements in the former equation:
i) $q_{i}^{2}=m_{\ell}^{2}$;
ii) $M_{23}^{2}=2(m_{\ell}^{2}+q_{2}\cdot q_{3})$; 
iii) $q_{2}\cdot q_{3}=\frac{1}{2}(M_P^{2}-4m_{\ell}^{2})-q_{1}\cdot q_{2}-q_{1}\cdot q_{3}-
q_{1}\cdot q_{4}-q_{2}\cdot q_{4}-q_{3}\cdot q_{4}$;
iv) $q_{1}\cdot q_{3}=\frac{1}{2}(M_{134}^{2}-3m_{\ell}^{2})-q_{1}\cdot q_{4}-q_{3}\cdot q_{4}$;
v) $q_{2}\cdot q_{4}=\frac{1}{2}(M_{124}^{2}-3m_{\ell}^{2})-q_{1}\cdot q_{2}-q_{1}\cdot q_{4}$;
vi) $q_{1}\cdot q_{2}=\frac{1}{2}M_{12}^{2}-m_{\ell}^{2}$, $q_{1}\cdot q_{4}=\frac{1}{2}M_{14}^{2}-m_{\ell}^{2}$, 
$q_{3}\cdot q_{4}=\frac{1}{2}M_{34}^{2}-m_{\ell}^{2}$.
Finally, the expression for the interference term in Eq.~(\ref{amplitude}) reads
\end{small}

\end{multicols}
\begin{eqnarray}
&&\frac{e^{8}|F(M_{12}^{2},M_{34}^{2})||F(M_{14}^{2},M_{23}^{2})|}{M_{12}^{2}M_{34}^{2}M_{14}^{2}
\left(2m_{\ell}^{2}+M_{P}^{2}-M_{124}^{2}-M_{134}^{2}+M_{14}^{2}\right)^{2}}\Big\{-2m_{\ell}^{8}+M_{12}^{6}\left(M_{134}^{2}-M_{14}^{2}\right)
+4m_{\ell}^{6}\left(-M_{12}^{2}+M_{124}^{2}+M_{134}^{2}-M_{34}^{2}\right)
\nonumber\\
&&+m_{\ell}^{4}\Big[M_{12}^{4}-3\left(M_{124}^{2}+M_{134}^{2}\right)^{2}+4M_{P}^{2}M_{14}^{2}+2M_{12}^{2}
\left(M_{124}^{2}+5M_{134}^{2}-2M_{14}^{2}-5M_{34}^{2}\right)+2\left(5M_{124}^{2}+M_{134}^{2}-2M_{14}^{2}\right)M_{34}^{2}+M_{34}^{4}\Big]
\nonumber\\
&&-\left(M_{124}^{4}+M_{134}^{4}-2M_{P}^{2}M_{14}^{2}\right.\left.-2M_{134}^{2}M_{34}^{2}+2M_{14}^{2}M_{34}^{2}+M_{34}^{4}\right)
\times\Big[M_{124}^{2}\left(M_{134}^{2}-M_{34}^{2}\right)+M_{14}^{2}\left(-M_{P}^{2}+M_{34}^{2}\right)\Big]
\nonumber\\
&&+M_{12}^{4}\Big[\left(M_{P}^{2}+2M_{134}^{2}-2M_{14}^{2}\right)M_{14}^{2}+\left(-2M_{134}^{2}+M_{14}^{2}\right)M_{34}^{2}+M_{124}^{2}
\left(-3M_{134}^{2}+2M_{14}^{2}+M_{34}^{2}\right)\Big]
\nonumber\\
&&+m_{\ell}^{2}\Big[-M_{12}^{6}+\left(M_{124}^{2}+M_{134}^{2}\right)^{3}-\left(M_{124}^{2}+M_{134}^{2}\right)
\times\left(5M_{124}^{2}+M_{134}^{2}-4M_{14}^{2}\right)M_{34}^{2}+\left(M_{124}^{2}+M_{134}^{2}-4M_{14}^{2}\right)M_{34}^{4}
\nonumber\\
&&-M_{34}^{6}+M_{12}^{4}\left(M_{124}^{2}+M_{134}^{2}-4M_{14}^{2}+M_{34}^{2}\right)+M_{12}^{2}\Big\{-\left(M_{124}^{2}+M_{134}^{2}\right)
\left(M_{124}^{2}+5M_{134}^{2}-4M_{14}^{2}\right)+2M_{P}^{2}\left(M_{124}^{2}-M_{134}^{2}+2M_{14}^{2}\right)
\nonumber\\
&&+2\Big[3\left(M_{124}^{2}+M_{134}^{2}\right)+4M_{14}^{2}\Big]M_{34}^{2}+M_{34}^{4}\Big\}-2M_{P}^{2}\Big[M_{124}^{4}
+\left(M_{134}^{2}+2M_{14}^{2}\right)\left(M_{134}^{2}-M_{34}^{2}\right)+M_{124}^{2}\left(-2M_{134}^{2}+2M_{14}^{2}+M_{34}^{2}\right)\Big]\Big]
\nonumber\\
&&+M_{12}^{2}\Big\{M_{134}^{6}+M_{124}^{4}\left(3M_{134}^{2}-M_{14}^{2}-2M_{34}^{2}\right)-M_{134}^{4}\left(M_{14}^{2}+2M_{34}^{2}\right)
-2M_{124}^{2}\left(M_{P}^{2}M_{14}^{2}\right.\left.+M_{134}^{2}M_{14}^{2}-M_{134}^{2}M_{34}^{2}+2M_{14}^{2}M_{34}^{2}+M_{34}^{4}\right)
\nonumber\\
&&+M_{134}^{2}\Big[-2M_{P}^{2}M_{14}^{2}+M_{34}^{2}\left(-4M_{14}^{2}+M_{34}^{2}\right)\Big]+M_{14}^{2}\Big[M_{34}^{2}
\left(4M_{14}^{2}+M_{34}^{2}\right)+M_{P}^{2}\left(4M_{14}^{2}+6M_{34}^{2}\right)\Big]\Big\}\Big\}\ .
\end{eqnarray}

\vspace{-1mm}
\centerline{\rule{80mm}{0.1pt}}
\vspace{2mm}

\begin{multicols}{2}

\end{multicols}

\clearpage

\end{CJK*}
\end{document}